\documentclass[twocolumn,prb,showpacs,preprintnumbers,amsmath,amssymb]{revtex4}
\usepackage{graphicx}
\usepackage{mathptmx}
\usepackage{verbatim}
\usepackage{graphicx}
\usepackage{amsmath}
\usepackage{amssymb}
\usepackage{mathdots}
\usepackage{amsfonts}
\usepackage{mathrsfs}
\usepackage{amsmath}
\usepackage{color}
\usepackage{graphicx}
\usepackage{bm}
\usepackage{amssymb}
\usepackage{xspace}
\usepackage{dcolumn}

\begin{document}

\title{Ab-Initio Theory of Moir{\' e} Superlattice Bands in Layered Two-Dimensional Materials}

\author{Jeil Jung}
\affiliation{The University of Texas at Austin, Austin, Texas 78712, USA }
\affiliation{National University of Singapore, 117551, Singapore.}
\author{Arnaud Raoux}
\affiliation{CNRS UMR 8502, Univ. Paris-Sud, F-91405 Orsay, France}

\author{Zhenhua Qiao} 
\affiliation{The University of Texas at Austin, Austin, Texas 78712, USA }

\author{A.H. MacDonald} 
\affiliation{The University of Texas at Austin, Austin, Texas 78712, USA }


\begin{abstract} 
When atomically thin two-dimensional (2D) materials are layered they often form 
incommensurate non-crystalline structures that exhibit long-period moir{\' e} patterns 
when examined by scanning probes.  
In this paper we present an approach which uses information obtained from
{\it ab initio} calculations performed on  short-period crystalline structures  
to derive effective Hamiltonians that are able to efficiently describe the influence 
of the moir{\' e} pattern superlattices on electronic properties.  We apply our approach to 
the cases of graphene on graphene (G/G) and graphene on hexagonal boron nitride (G/BN), deriving explicit effective  
Hamiltonians that have the periodicity of the moir{\' e} pattern
and can be used to calculate electronic properties of interest for 
arbitrary twist angles and lattice constants.
\end{abstract}


\maketitle

\section{Introduction}

Since shortly after it was first isolated for electronic property studies in 2004,~\cite{Novoselov04}
the graphene family of two-dimensional electron systems has attracted great interest.
Recently attention has expanded\cite{NovoselovNobel} to include other extremely anisotropic
materials, including hexagonal boron nitride\cite{Dean10} (hBN)
and transition metal dichalcogenides,\cite{TMDs} and to structures in which 
combinations of these materials are stacked in various different ways.  
All these materials share hexagonal 
lattice structures, and have low-energy electronic states located  
at momenta near the two-dimensional lattice's Brillouin-zone corners.  


Because the lattice constants of these 2D materials differ, and because 
the hexagonal lattice orientations of different layers are not always identical, 
multilayer systems usually do not form two-dimensional crystals.
For example, the lattice constant of hBN is approximately   
$1.7\%$ larger than that of graphene.  Differences in lattice constant or orientation produce 
moir{\' e} patterns\cite{moire_graphite} that are apparent in scanning probe studies 
of electronic properties\cite{Pong05,Amidror09,Andrei,LeRoy} when graphene is placed on a graphite or hBN substrate.
The moir{\' e} pattern is responsible for Hofstadter\cite{HofstaderTheoryOld} 
gaps\cite{HofstaderExptOld,HofstaderExpt,chou,HofstaderTheory} that occur 
within Landau levels when samples are placed in a perpendicular magnetic field.   

The period of the moir{\' e} pattern is unrelated to true two-dimensional
crystallinity, which for a given lattice constant 
difference is present only at discrete relative orientations, and appears to have little 
relevance for observed properties.  The absence of crystallinity nevertheless complicates theoretical 
descriptions of electronic properties because it removes
the simplifications which would otherwise be afforded by Bloch's theorem.  
This obstacle has forced researchers to proceed either 
by using simplified tight-binding models,\cite{Santos07,Mayou,Mele,Kindermann,RafiTun} or by performing 
{\it ab initio} or tight-binding calculations\cite{Shallcross,linhe,chou} for long-period crystal approximants to real structures.   
An alternative approach,\cite{RafiMoireBand,SantosII,montambaux2011,sanjose3,sanjose1,sanjose2,gonzalez,calvin,moonkoshino} is 
based on the assumption that interlayer tunneling amplitudes in 2D materials vary slowly on an atomic scale with changes in 
either initial or final two-dimensional position.  When this assumption is valid, 
it is possible to formulate an effective theory of low-energy electronic structure in which the 
Hamiltonian is periodic with the periodicity of the moir{\' e} pattern, and therefore to employ Bloch's theorem.  
We refer to models of this type, which seek mainly to describe electronic properties
in systems with long moir{\' e} period structures over a limited energy range, as moir{\'e} band models.  

In this paper we extend the moir{\' e} band approach,
explaining how moir{\' e} band models can be systematically obtained from 
{\em ab initio} electronic structure calculations performed only on short-period 
commensurate multilayer structures.  
The moir{\' e} band Hamiltonian is position dependent and acts
on real spin and on orbital, sublattice and layer pseudo spin degrees of freedom.
A moir{\' e} band Hamiltonian does not account for the 
presence or absence of commensurability between the 
underlying lattices.  Moir{\' e} band Hamiltonians are particularly advantageous for 
theories of electronic properties in the presence 
of an external magnetic field for which other more direct approaches are usually not practical.  
We illustrate our method of deriving moir{\' e} band Hamiltonians 
by applying it to the case of two graphene layers, and to the case of graphene on hBN.

Our paper is organized as follows.  In Section ~\ref{sec:formal} we explain our 
approach, which can be applied to any system of layered 2D materials in which the local inter-layer 
stacking arrangement varies slowly on an atomic scale.  The parameters of the 
model can be extracted from {\em ab initio} electronic structure 
calculations by examining the dependence of electronic states on 
relative displacements between layers in crystalline stacked structures.  
In Section  ~\ref{sec:abinitio} we discuss the 2D band structures of crystalline 
graphene on graphene (G/G) \cite{SantosII,Mayou,Mele,Kindermann,RafiTun,Shallcross,RafiMoireBand,montambaux2011,sanjose1,sanjose2,sanjose3,gonzalez,linhe,calvin,moonkoshino}
and graphene on hexagonal boron nitride (G/BN).  \cite{giovannetti07,sachs,slawinska,bokdam,Arnaud,falko1,falko2} 
We extract the quantitative values of the small number of parameters 
which characterize the corresponding moir{\' e} band models from these calculations.
(As experimental results emerge, the model parameters could instead 
be fit to observed properties if deemed more reliable than {\em ab initio} 
electronic structure calculations.)  
In Sections ~\ref{sec:grapheneongraphene} and ~\ref{sec:grapheneonbn} we 
describe applications to G/G and G/BN.  
For G/BN, it is possible to 
derive a two-band moir{\' e} band model which describes only the graphene layer $\pi$-bands.
We expect that this simplified model, which is explained in Section 
~\ref{sec:twoband}, will be widely applicable to evaluate many physical properties of 
graphene on BN substrates.  
Finally in Section ~\ref{sec:discussion} we summarize our results and briefly 
discuss some issues which may arise in applying our approach to other 2D material stacks, 
and in accounting many-body effects that are important for theories of 
some physical properties.

\section{Moir{\' e} Band Model Derivation}
\label{sec:formal}

Our method applies to stacks of two-dimensional crystals with the same lattice structure,
similar lattice constants, and relative orientation angles that are not too large.
It is ideally suited to stacks composed of graphene and hBN layers in arbitrary order 
with arbitrary orientations, or to group VIB transition-metal dichalcogenide semiconductor stacks.
The basic idea is that the lattice representation Hamiltonian,
\begin{equation}
\label{Hlat}
{ H}_ {lat} = \langle l s {\vec{L}} | { H} | l' s' \vec{L}' \rangle,
\end{equation} 
depends mainly 
on the local coordination between layers $l$ and $l'$, and that this dependence can
be characterized performing calculations for crystalline structures in which 
the layers are displaced arbitrarily, but share the same lattice constant and orientation.  
In Eq.~(\ref{Hlat}) $l$ labels the layers, each of which is assumed to form a 2D crystal, 
$s$ labels sites within the 2D crystal unit cell, and $\vec{L}$ labels lattice vectors.  
If more than one atomic orbital were relevant at each lattice site, as would be the case for
transition metal dichalcogenides for example, $s$ would label both site and relevant orbitals on that site.  
For graphene and hBN we will restrict our attention to the $\pi$-bands so we will consider only
one orbital per atom.  

The moir{\' e} band model is defined by matrix elements of ${ H}_{lat}$ 
calculated in the representation of the 2D Bloch states of the individual layers.
Below we first explain the approximation we use for ${ H}_{lat}$, and then
explain how we use it to evaluate Bloch state matrix elements.  We will focus on 
the case of 2D honeycomb lattices so that our discussion applies specifically to 
the graphene and hBN cases of primary interest.  
In this paper we focus on the two layer case, and comment on the more general case only in the 
discussion section.  

When the individual layer 2D lattices have the same orientation and identical lattice constants, the overall 
material is crystalline.  In that case we can exploit translational symmetry and
solve the electronic structure problem using Bloch's theorem.  
Using Bloch state completeness properties it is easy to show that 
\begin{widetext} 
\begin{equation} 
\langle l s {\vec {L}} | { H}(\vec{d}) | l' s' \vec{L}' \rangle = 
\frac{1}{N} \sum_{\vec{k} \in BZ} \exp(i \vec{k} \cdot (\vec{L}+\vec{\tau}_{s}))  \; { H}_{ls,l's'}(\vec{k}:\vec{d}) \;
\exp(-i \vec{k} \cdot (\vec{L}'+\vec{\tau}_s')).
\label{crystal}
 \end{equation}
Here $ { H}_{ls,l's'}(\vec{k}:\vec{d})$ is the Wannier representation Bloch-band Hamiltonian;
we have explicitly indicated that it is a non-trivial function of any rigid displacement $\vec{d}$ of the top layer 
with respect to the bottom layer.  (We include the displacement in the site positions so that 
i.e. $\vec{\tau}_{s'} \to \vec{\tau}_{s'} + \vec{d}$ in the top layer when it is displaced.
$\vec{d}$ is defined to be zero for AA stacking.)
Note that $\langle ls \vec{L} | { H}(\vec{d})| l' s' \vec{L}' \rangle$ is a function 
only of $\vec{L}'-\vec{L}$, and not of $\vec{L}$ and $\vec{L}'$ separately, and that
${ H}_{ls,l's'}(\vec{k}:\vec{d}) = { H}_{ls,l's'}(\vec{k}:\vec{d}+\vec{L})$.
The geometry of two stacked honeycomb lattices is illustrated in Fig.~\ref{schematic}.  

\begin{figure*}[htbp]
\includegraphics[width=17cm,angle=0]{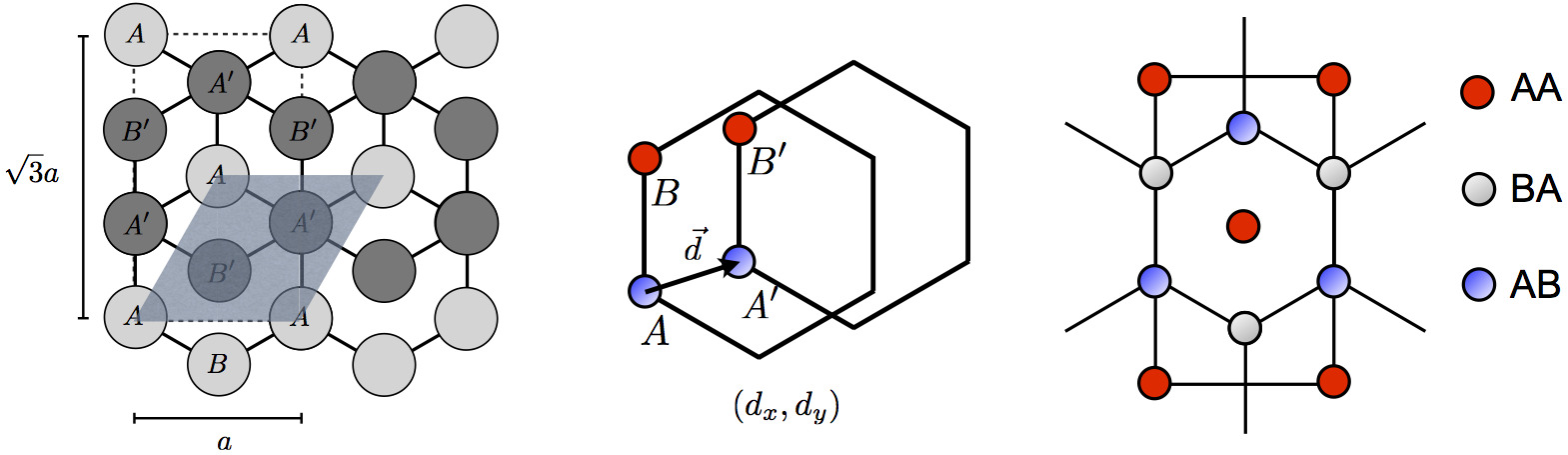}
\caption{(Color online)
Left: Schematic representation of two commensurate honeycomb layers with 
bottom layer sites indicated by light grey circles and top layer sites indicated by 
dark grey circles.  The unit cell of the bilayer contains four sites, $A$ and $B$ for bottom layer 
and $A'$ and $B'$ for the top layer. The shaded region represents
the primitive cell area $A_0$ used for the Fourier integrals described in the text.  
Middle: The relative displacement between the 
honeycombs is specified by the displacement vector $\vec{d}$.  We choose 
$\vec{d}=0$ for AA stacking in which the two honeycombs have no lateral displacement.  
For $\vec{d} = (0, a / \sqrt{3})$ we have the AB stacking where the top layer $A'$ site is directly
above the bottom layer $B$ site.
The bilayer lattice is a periodic function of $\vec{d}$ and the primitive cell 
for this periodic dependence is shaded grey in the left figure.  It is 
convenient  to use the rectangular $a \times \sqrt{3}a$ area enclosed by a dotted line in the left figure 
to illustrate the dependence of the bilayer 
Bloch bands on $\vec{d}$.  Right: This panel specifically indicates the 
points within the rectangular area at which 
the high symmetry $AA$, $BA$, and $AB$ stacking arrangements occur and 
is helpful for the interpretation of later figures.}
\label{schematic}
\end{figure*}

The moir{\' e} band model is intended to provide a low-energy effective model of electronic 
states for the case in which the top layer lattice is expanded by factor
$\alpha$ and rotated counterclockwise by rotation rotation angle $\theta$ 
with respect to the bottom layer lattice.
Note that rigid displacements of incommensurate layers lead only to a 
spatial shift in the moir{\' e} pattern and otherwise have no effect on the electronic structure. \cite{RafiMoireBand}
When we wish to retain the dependence on initial translation,
we denote it by $\vec{\tau}$ . 
For simplicity we first discuss the case in which $\vec{\tau} = 0$,
and later restore the matrix-element phase-factor changes introduced by this initial translation.
The shift in lattice 
positions of the top layer with respect to the original positions 
can then be expressed in terms of $\alpha$ and the rotation operator $\cal R$$(\theta)$:
\begin{equation} 
\label{dofL}
\vec{d}(\vec{L}) \equiv \alpha {\cal{R}}(\theta) \vec{L} - \vec{L} = 
\left( (\alpha \cos(\theta) - 1)L_x - \alpha \sin(\theta) L_y, (\alpha \cos(\theta)-1) L_y + \alpha \sin(\theta) L_x \right).
\end{equation} 
\end{widetext} 
We obtain our moir{\' e} band model by approximating
the lattice matrix elements of the scaled and rotated structure 
using Eq.~(\ref{crystal}) with $\vec{d}$ replaced by $\vec{d}(\vec{L})$ in Eq.~(\ref{dofL}).
In using this approximation we are assuming that 
$\vec{d}(\vec{L})$ varies slowly on an atomic length scale, {\em i.e.} that 
$\theta$ and $\alpha-1 \equiv \epsilon$ are small.  
In this limit 
\begin{equation} 
\vec{d}(\vec{L}) = \epsilon \, \vec{L} \, + \,  \theta \, \hat{z} \times \vec{L}. 
\end{equation} 
(The distinction between $\vec{d}(\vec{L})$ and $\vec{d}(\vec{L'})$ is second order 
in the small parameters $\epsilon, \theta$ and therefore neglected.
The moir{\' e} pattern formed by rotation and scaling is discussed further in Appendix \ref{exactform}.)    
Once this substitution is made $\langle ls \vec{L} | { H}(\vec{d})| l's' \vec{L}' \rangle$ 
depends on both $\vec{L}$ and $\vec{L}'$ and not just on $\vec{L}' - \vec{L}$. 
It follows that the Hamiltonian is no 
longer Bloch diagonal in a momentum space representation.  Our local displacement approximation 
is obviously exact for $\epsilon=\theta=0$, and we believe that it is accurate 
over useful ranges of $\alpha$ and $\theta$ as discussed further below.  
We defer further comment on the accuracy of the 
approximation until we discuss the two explicit examples explored in this article,
G/G and G/BN.

In order to use this approximation conveniently, we 
note that ${ H}_{ls,l's'}(\vec{k}:\vec{d})$  is a periodic function of 
$\vec{d}$ with lattice periodicity, {\em i.e.} that ${ H}_{ls,l's'}(\vec{k}:\vec{d}) = { H}_{ls,l's'}(\vec{k}:\vec{d}+\vec{L})$.  
It can therefore be expanded in terms of reciprocal lattice vectors
\begin{eqnarray} 
{ H}_{ls,l's'}(\vec{k}:\vec{d}) &=& \sum_{\vec{G}} { H}_{ls,l's'}(\vec{k}:\vec{G})
\, \exp( - i \vec{G} \cdot \vec{d} )     \nonumber     \\
&\times& \exp( -i \vec{G} (\vec{\tau}_{s'} - \vec{\tau}_{s}) \widetilde{\delta}_{ll'}).
\label{Hofd} 
\end{eqnarray} 
The phase factor $\exp( -i \vec{G} (\vec{\tau}_{s'} - \vec{\tau}_{s}) \widetilde{\delta}_{ll'})$
is included in the definition of the Fourier expansion coefficients in order to make 
their symmetry properties more apparent, and $\widetilde{\delta}_{ll'} = (1 - \delta_{ll'})$ where $l, l'$ are layer indices.
We show below that for G/G and G/BN
only a few terms in this Fourier expansion are large.  As we explain there, we expect this to 
be a general property of 2D material stacks.  

${ H}_{ls,l's'}(\vec{k}:\vec{d})$ can be calculated
relatively easily by performing {\em ab initio} 
supercell  density-functional-theory (DFT).  The number of atoms per unit 
cell in these calculations is modest, four for example in the cases with two crystal layers with two atoms per cell
considered explicitly in this paper.  
The Fourier coefficients which describe the dependence of ${ H}_{ls,l's'}(\vec{k}:\vec{d})$
on $\vec{d}$ are obtained 
by evaluating the inverse Fourier transform numerically: 
\begin{eqnarray} 
{ H}_{ls,l's'}(\vec{k}:\vec{G}) &=& \frac{1}{A_0} \int_{A_0} d \vec{d}\; 
 { H}_{ls,l's'}(\vec{k}:\vec{d})    \exp(i \vec{G} \cdot \vec{d})  \nonumber  \\
 &\times& \exp( i \vec{G} (\vec{\tau}_{s'} - \vec{\tau}_{s}) \widetilde{\delta}_{ll'})
\label{integrateoverd}
\end{eqnarray} 
where $A_0$ is the integration area of a commensurate configuration primitive cell
shown in Fig. \ref{schematic}.

\begin{figure*} 
\begin{center}
\begin{tabular}{cc}
\includegraphics[width=8cm,angle=0]{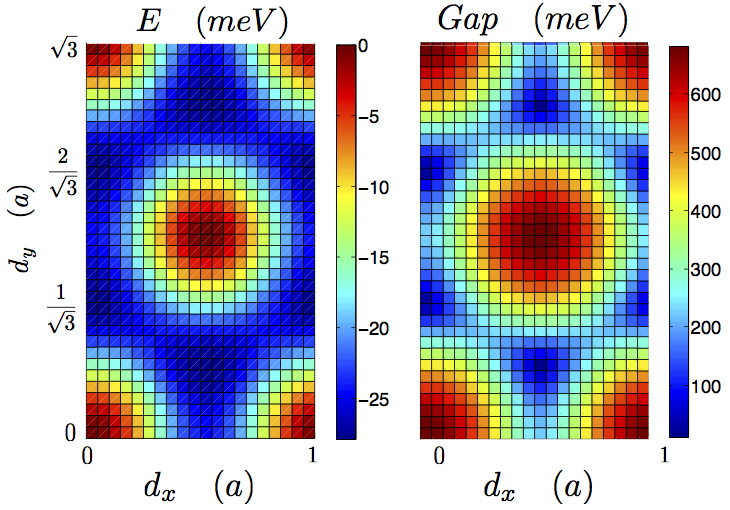}  \quad \,\, & \,\, \quad 
\includegraphics[width=8cm,angle=0]{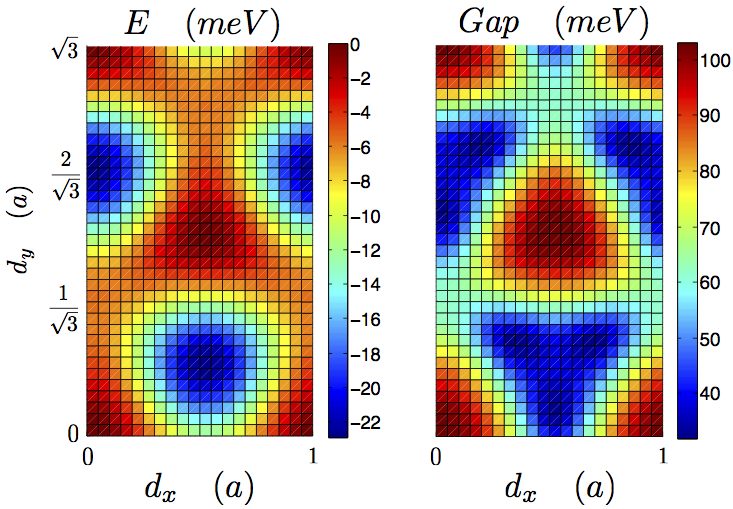}
\end{tabular}
\end{center}
\caption{(Color online)
{\em Left panel:}
Total energy per unit cell relative to AA stacking as a function of displacement $\vec{d}$. 
These results are for G/G with 
constant in-plane lattice constant and vertical separation $c=3.35~\AA$.
The highest energy configuration corresponds to AA stacking and the lowest 
to AB or BA stacking. {\em Gap} refers to the separation between conduction and 
valence bands at the Dirac point, which vanishes at AB and BA points.  
{\em Right panel:}
The same plots for G/BN with both sheets constrained to have the in-plane
self-consistent LDA lattice constant of graphene and vertical separation $c = 3.35~\AA$.  
The lowest energy stacking configuration corresponds to BA stacking with one of the 
two carbon atoms in the graphene unit cell sitting on top of boron.  The highest energy stacking
configuration corresponds to the AA arrangement 
in which the two carbon atoms in the unit cell sit on top 
of B and N.  The AB configuration which has C on top of N has an intermediate energy.  
The scale of the dependence of total energy on $\vec{d}$
is similar in the graphene and hBN cases.  
Note that the Dirac point gap in the G/hBN case does not vanish at any value of $\vec{d}$,
but that the typical gap scale is larger in the G/G case.  This later property reflects
stronger inter-layer coupling.   
 }
 \label{slidingfig_gg}
\end{figure*}

We are now in a position to derive our low-energy model.  
First of all we use Eqs. (\ref{dofL}) and ~(\ref{Hofd}) 
to construct the momentum space matrix elements of our model.  We assume that each layer is 
still accurately crystalline and for each layer evaluate matrix elements using Bloch states defined using the 
two-dimensional crystal structure of that layer.  Summing independently over the lattice vectors $\vec{L}'$ of layer $l'$ and 
$\vec{L}$ of layer $l$ and using Eqs. ~(\ref{crystal}), (\ref{dofL}), and ~(\ref{Hofd}) leads after an elementary calculation to 
\begin{widetext} 
\begin{equation} 
\langle l s \vec{k} | { H} | l' s' \vec{k'} \rangle = 
\sum_{G} H_{ls,l's'}(\vec{k}:\vec{G}) \; \exp(i\vec{G} \cdot (\vec{\tau}_{s} - \vec{\tau}_{s'}) \widetilde{\delta}_{ll'}) \; 
\Delta(\vec{k'}-\vec{k} - \tilde{ G}) 
\label{matrixelement}
\end{equation} 
\end{widetext} 
where $\Delta(\vec{k})=1$ when $\vec{k}$ is a reciprocal lattice vector and is zero otherwise and 
\begin{eqnarray}
\widetilde{ G} &=&  \epsilon \vec{G} -\theta \, \hat{z} \times \vec {G}.   
\label{tildeG}
\end{eqnarray}

In applying this formula we wish to describe electronic states 
derived from Bloch orbitals close to a particular or several particular points in momentum space.
For graphene and hBN we wish to describe states close to the 
Dirac points $\vec{K}$ and $\vec{K}'$.  (Below we consider the $\vec{K}$ Dirac 
point for definiteness.)   
We further assume that the interlayer coupling processes responsible for $\l \ne l'$ and 
$\vec{G} \ne 0$ terms in Eq.~(\ref{matrixelement}) are small compared to the $\l = \l'$, $\vec{G} = 0$ 
term, but that they vary with momentum $\vec{k}$ on the same reciprocal lattice vector scale.
These assumptions allow us to replace $H_{ls,l's'}(\vec{k}:\vec{G})$ by its value at the
Dirac point when $l' \ne l$ or $\vec{G} \ne 0$ to obtain  
\begin{widetext} 
\begin{eqnarray} 
\langle l s \vec{k} | { H} | l' s' \vec{k'} \rangle &=& \delta_{l,l'}
\big[ H_{ls,l's'}(\vec{k}:\vec{G}=0)  \delta_{\vec{k},\vec{k}'}  + \sum_{\vec{G}\ne0}  H_{ls,l's'}(\vec{K}:\vec{G}) 
\; \Delta(\vec{k'}-\vec{k} - \tilde{ G})  \big] \nonumber \\
&+&  \widetilde{\delta}_{l,l'} \sum_{\vec{G}} H_{ls,l's'}(\vec{K}:\vec{G}) \exp(i\vec{G} \cdot (\vec{\tau}_{s} - \vec{\tau}_{s'})) \; 
\Delta(\vec{k'}-\vec{k} - \tilde{ G})
\label{matrixelementbb}
\end{eqnarray} 
\end{widetext} 
where the first and second lines are respectively the intralayer and interlayer terms.

As we see in Eq.~(\ref{matrixelementbb}) the Hamiltonian is constructed as the sum of several 
contributions: i)~an isolated layer two-dimensional band structure obtained by averaging over 
displacements $\vec{d}$, ii)~a sublattice pseudo-spin dependent term which acts within layers and accounts for 
the influence of nearby layers on-site-energies and hopping within layers and iii)~an inter-layer 
tunneling term which is also strongly dependent on the local stacking arrangement.  
The first term in this equation reduces to the isolated layer Hamiltonian when interlayer 
coupling is absent.  In fact, according to our numerical calculations, the difference between this 
term and the isolated layer Hamiltonian is negligible for both G/G and G/BN cases.
The second and third terms are off diagonal in momentum and therefore account for the 
moir{\' e} pattern which breaks translational symmetry.
As we illustrate below, using G/G and G/BN as examples, useful models can be 
obtained with a small number of independent $H_{ls,l's'}(\vec{K}:\vec{G})$ parameters, partly because of  
symmetry.  The values of the parameters can be evaluated 
by using {\em ab initio} electronic structure calculations (the approach we follow in this paper) 
to examine the relative displacement $\vec{d}$ dependence of the electronic 
structure of two-dimensional layers that have a common Bravais lattice. 
Since the basic premises of DFT theory are reasonably reliable for graphene and hBN, 
the resulting model is expected to capture all qualitative electronic structure 
features associated with the moir{\' e} pattern.  More accurate simple models might eventually
be achievable by using this approach to construct similar phenomenological models 
with parameters derived from experimental observations.   
We illustrate the power of this simple formula below by applying it to the G/G and G/BN cases.  

\section{Ab-Initio Moir{\' e}-Band-Models} 
\label{sec:abinitio} 

\subsection{Electronic Structure Calculations} 

We calculate our moir{\' e} band parameters 
starting from Wannier-function 
lattice representations of bilayer perfect crystal Hamiltonians.  
In this section we present a brief summary of the first-principles
methods employed to obtain the Wannier-function
representation Hamiltonian matrices, and discuss some
qualitative aspects of the perfect crystal bands of G/G and G/BN
that hint at important moir{\' e} band properties.
Our microscopic calculations were performed for two-layer systems 
with four atoms per unit cell.  
We used the software package
{\em Quantum Espresso} \cite{espresso} that is interfaced with 
the package {\em wannier90}. \cite{wannier90,marzarirmp}
The calculations were performed using a $42 \times 42$ $k$-point sampling density,
an energy cutoff of 60~Ry, vonBarth-Car norm conserving pseudopotentials,
and the Perdew-Zunger LDA parametrization.
(C,B,N.pz-vbc.UPF)  The same $k$-point sampling
density was maintained for the Wannier representation construction
of the Hamiltonian projected to 10 localized orbitals, 6 corresponding to the $\sigma$
bonds and 4 to $p_z$ orbitals centered on the four atoms.  
The convergence criteria used for self-consistent 
total energy in the DFT 
calculations was $10^{-9}$~eV per unit cell.

Although mirror symmetry is broken in bilayers for general $\vec{d}$,
the coupling between $\pi$ and $\sigma$ bands is always weak because 
their energy separation near the Dirac point is $\sim 10$ eV, and large compared
to coupling matrix elements that are always smaller than 0.1 eV.\cite{bilayertb}
We therefore retain only the $\pi$-electron degrees of freedom
in our moir{\' e} band models.  Because there is only one $p_z$ orbital 
per carbon atom, the Wannier-representation Hamiltonians discussed 
below are $4 \times 4$ matrices with row and column indices that 
can be labelled by the four sites in a two-layer crystal.     
We characterized the dependence on the relative displacement 
between the layers by performing calculations on a 
$\vec{d}$-sampling grid with $21 \times 36$ points in the $a \times \sqrt{3}a$ area plotted in 
Fig. \ref{slidingfig_gg}.

We have chosen a coordinate system in which graphene's 
triangular Bravais lattice has primitive lattice vectors
\begin{equation}
\vec{a}_1=a(1,0),   \qquad\qquad     \vec{a}_2=a\Big(-{1 \over 2},{\sqrt{3} \over 2}\Big),
\label{primitivecell}
\end{equation}
where $a = 2.46{~\AA}$ is the lattice constant of graphene.
The corresponding primitive reciprocal lattice vectors are
\begin{equation}
\vec{b}_1 = \frac{2 \pi}{ a}(1,\frac{1}{ \sqrt{3} }) \ , \qquad\qquad 
\vec{b}_2=\frac{2 \pi}{a} \Big( 0, {2 \over \sqrt{3} } \Big).
\label{reciprocal}
\end{equation}
The $A$ and $B$ sublattice positions in the bottom layer are 
\begin{equation}
\vec{\tau}_{A} =  (0, 0, 0) , \, \quad 
\vec{\tau}_{B} =  (0, \frac{a}{\sqrt{3}} , 0).
\end{equation}
and the $A^{\prime}$ and $B^{\prime}$ positions in the second layer are
\begin{equation}
\label{A'B'}
\vec{\tau}_{A^{\prime}} =  ( d_x,  d_y, c) , \, \quad 
\vec{\tau}_{B^{\prime}} =  ( d_x, \frac{a}{\sqrt{3}} + d_y, c).
\end{equation}
In Eq.~(\ref{A'B'}) $c$ is the layer separation that we assume to be constant.  
Results for geometries where the out of plane $z$ direction coordinate 
is relaxed as function of $\vec{d}$ are discussed in Appendix ~\ref{relaxed}.

In Fig. ~\ref{slidingfig_gg} we plot results for 
the dependence of total energy and the  
band gap at the Dirac point on displacement $\vec{d}$ for both G/G and G/BN.
In the G/BN case, the difference in lattice constant between
graphene and hBN layers will play an essential role.
The bulk lattice constant of graphite is $a_{G} = 2.461~\AA$ 
whereas $a_{hBN} = 2.504~\AA$, implying a difference of about 1.7$\%$.
For the commensurate calculations summarized in Fig. ~\ref{slidingfig_gg}
we used the self-consistent LDA lattice constant 
of single layer graphene $a_G = 2.439~\AA$ for both graphene and boron nitride sheets. 
Notice that for G/BN there is a gap at the Dirac point 
at any value of $\vec{d}$.

\begin{figure} 
\begin{tabular}{c}
\includegraphics[width=8.4cm,angle=0]{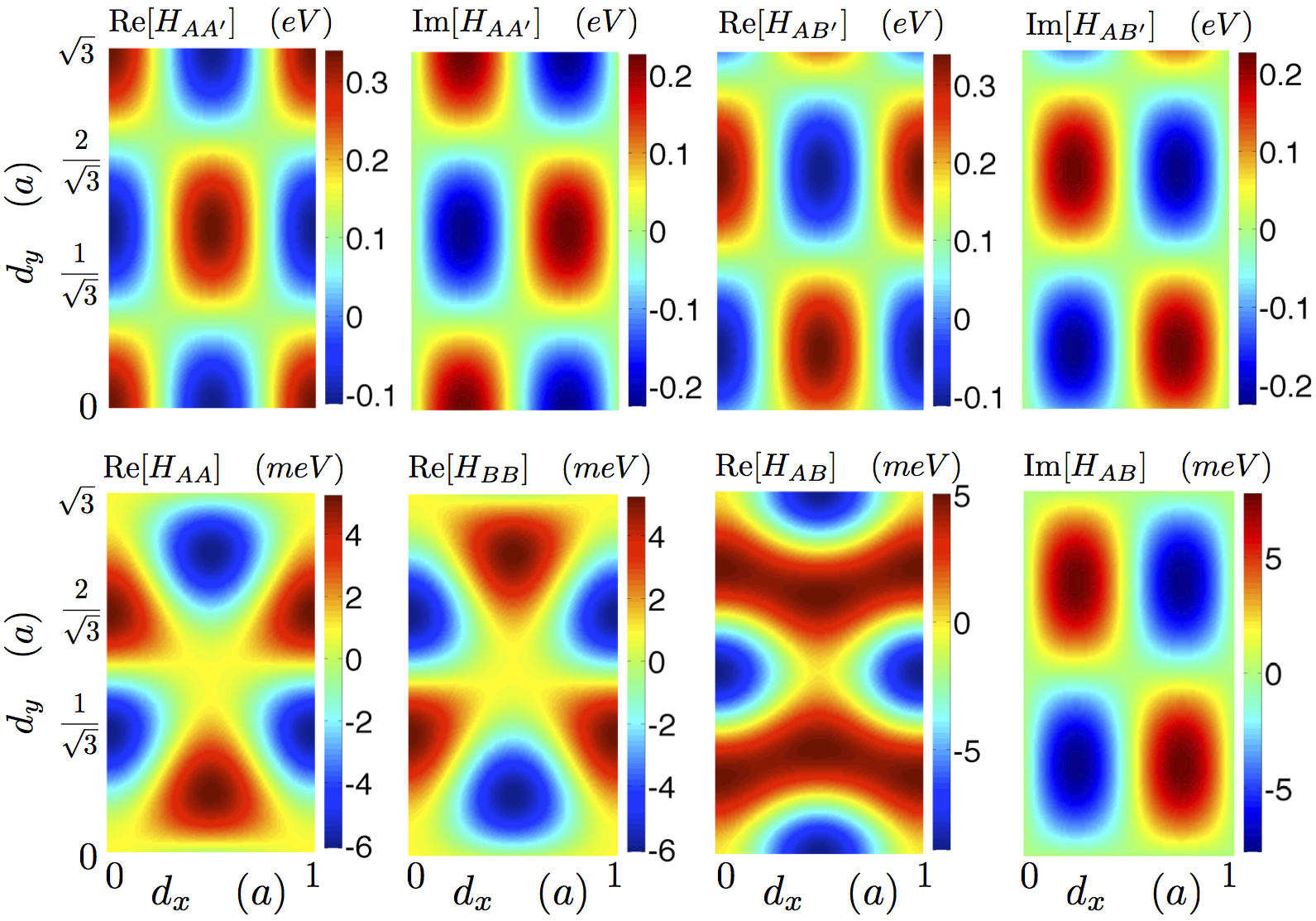} 
\end{tabular}
\caption{
(Color online)
Dirac-point $\pi$-band Wannier-representation Hamiltonian matrix
elements as a function of sliding vector $\vec{d}$ for graphene/graphene.
{\em Top panel:}
Left to right real and then imaginary parts of the $AA'$ and then $AB'$ interlayer matrix elements 
as a function of position $\vec{d}$ in the rectangular cell of Fig.1.  
The $BA'$ matrix element is closely related to the $AB'$ matrix element 
as shown in Eq.~(\ref{gginterlayer}) and the 
$BB'$ matrix element is identical to the $AA'$ matrix element.
Interlayer coupling matrix elements have a typical magnitude
$\sim$ 300 meV.  We show later that the dependence of these four 
complex numbers on $\vec{d}$ is accurately described by a single real number.
The color scales show energies in units of eV.
{\em Bottom panel:}
Intralayer Wannier-representation Hamiltonian matrix elements.  
Left to right the real parts of the $AA$ and $BB$ matrix elements followed 
by the real then imaginary parts of the $AB$ matrix element.
Typical matrix element values are $\sim 5$ meV.  
For graphene on graphene the spatial variation of intra-layer matrix 
elements has a negligible influence on electronic properties.
The color scale shows energy in units of meV.   
}
\label{interlayer1}
\end{figure}

\subsection{Graphene on Graphene} 
\label{sec:grapheneongraphene}
\subsubsection{Moir{\' e} Band Model} 

In Fig. \ref{interlayer1} we illustrate the dependence on $\vec{d}$ of
both interlayer and intra-layer values of $H_{ls,l's'}(\vec{K}:\vec{d})$ for the case of two-coupled graphene 
layers.  The intra-layer parameters are typically 
$\sim 5$ meV in the graphene case and do not play an essential role in the 
moir{\' e} bands; we will see later that their role is much
more essential in the graphene on boron nitride case.  
The inter-layer coupling at the Dirac point is larger 
and more strongly dependent on $\vec{d}$.  As we now 
explain, a single real parameter is sufficient to accurately describe the 
full $\vec{d}$ dependence of the four complex inter-layer coupling matrix 
elements.  The vast simplification is related to the smooth variation of 
inter-layer coupling on $\vec{d}$, which is related\cite{RafiMoireBand} in turn to the fact that the 
distance between layers of these van der Waals coupled two-dimensional materials
is substantially larger than the separation between atoms within a layer.  

\begin{figure}[htbp]
\label{gg_model_parameters}
\begin{tabular}{c}
\includegraphics[width=8cm,angle=0]{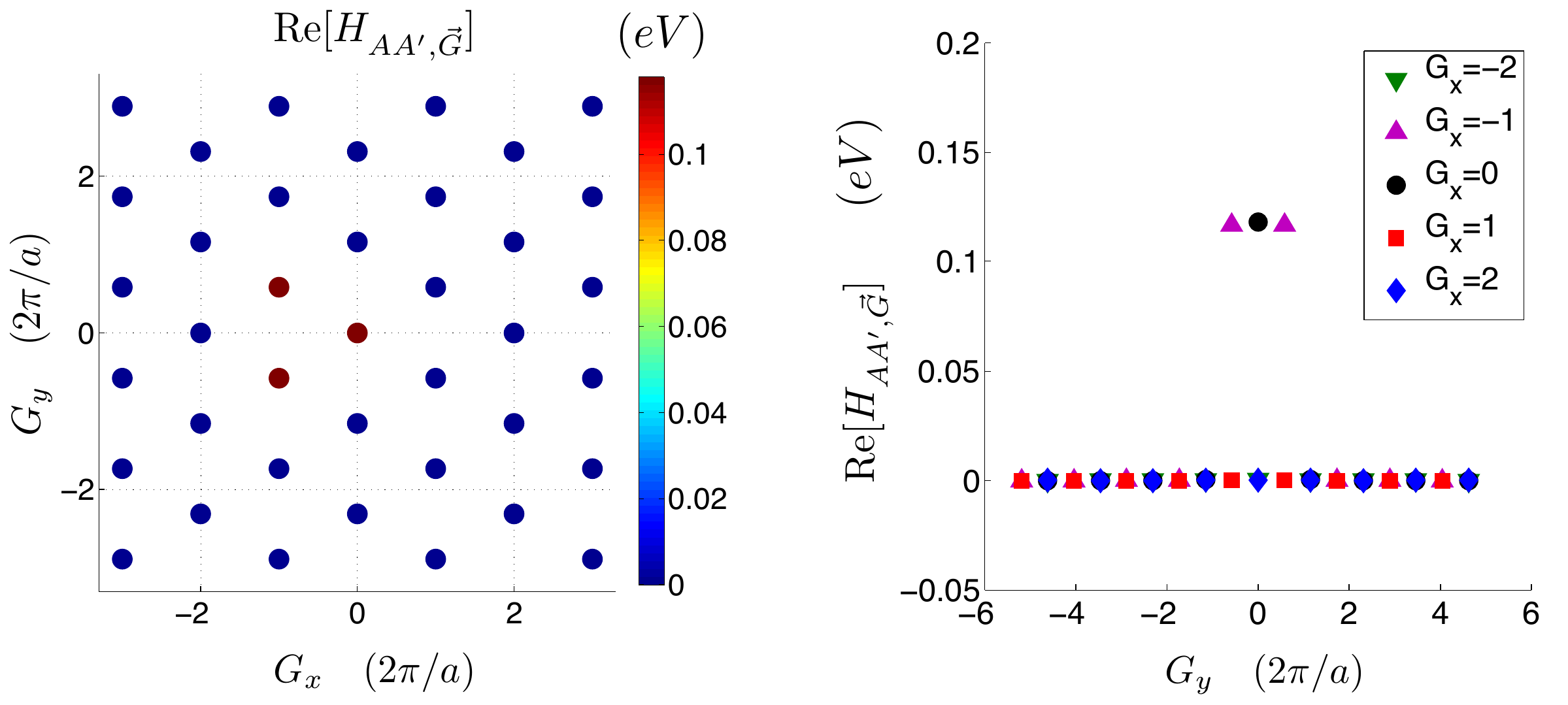} 
\end{tabular}
\caption{
(Color online)
Real part of the Fourier transform of $H_{AA'}(\vec{K}: \vec{d})$ evaluated at $\vec{K} = (4 \pi/3a,0)$.
In Fourier space, interlayer coupling is strong only for three
reciprocal lattice vectors, $\vec{G}=0$ and the 
two non-zero reciprocal lattice vectors for which $|\vec{K} + \vec{G}|=|\vec{K}|$.   
The imaginary part of $H_{AA'}( \vec{K} :{\vec{d}})$ vanishes.
At these values of $\vec{G}$, $H_{AA'}( \vec{K}: \vec{ G})$ is real with
identical values of $0.113 \pm 0.001$ eV.   
}
\end{figure}

Fig. 4 illustrates typical results of a moir{\' e} band parameter 
calculation performed by using Eq.~(\ref{integrateoverd})  and integrating over $\vec{d}$.  
We find that for graphene on graphene the only large corrections to the isolated layer Hamiltonian are 
for inter-layer tunneling, and that these are large at $\vec{G}=0$ and at two 
non-zero values of $\vec{G}$, that they are real, and that they are identical in 
the three cases.  The interlayer parameters do not have 
the symmetry of the reciprocal lattice because they are evaluated at the 
Brillouin-zone corner Dirac point, rather than at the zone center. 
The three $\vec{G}$'s which yield large parameters share the 
minimum value of $|\vec{K} + \vec{G}|$.
The entire inter-layer coupling part of the Hamiltonian is accurately captured by 
a single real parameter with the value $0.113 \pm 0.001$ eV. 
We now explain the physics behind this seemingly surprising simplification.  

\begin{figure}
\includegraphics[width=8cm]{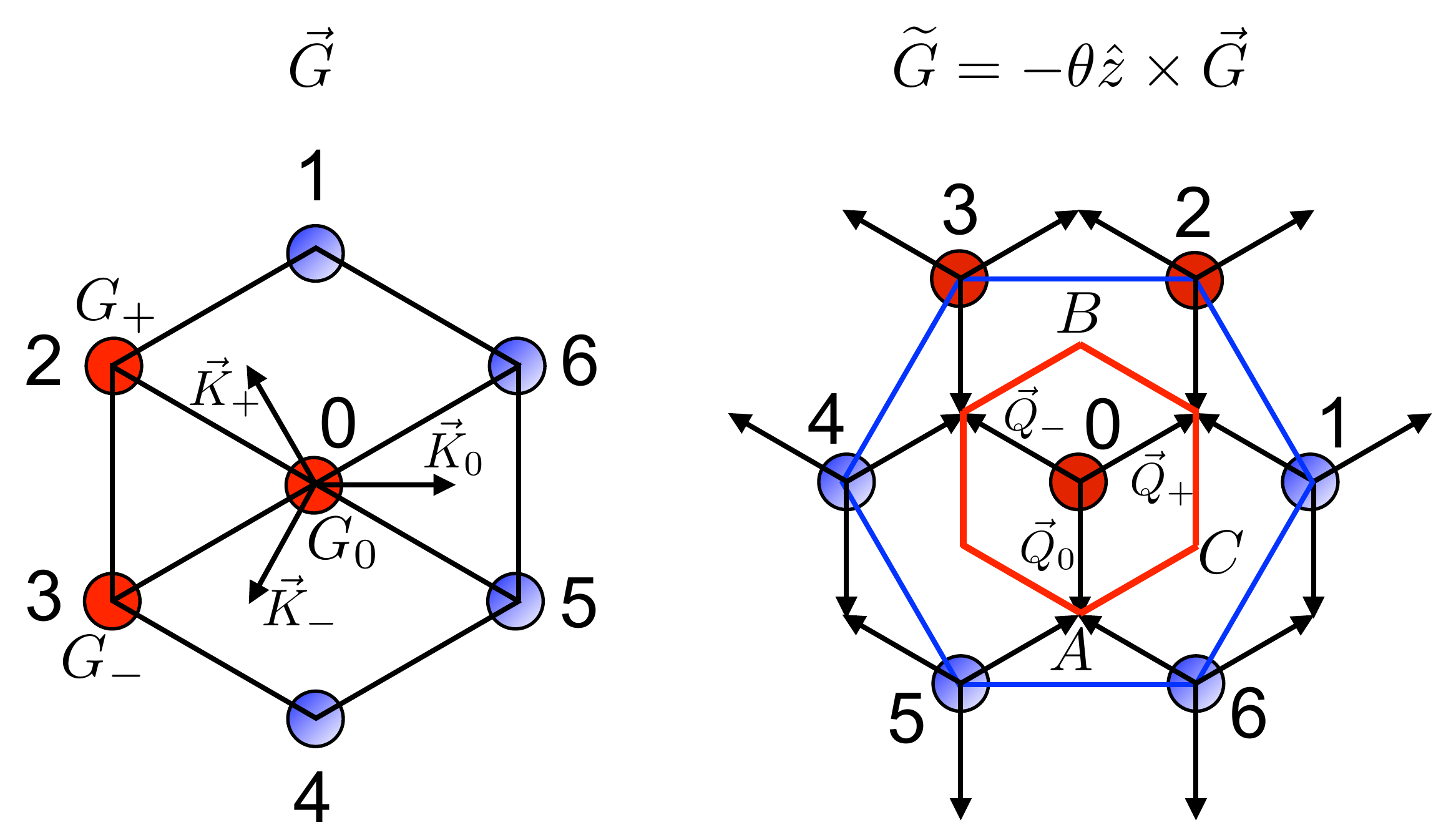} \\
\caption{
{\em Left Panel:} Representation of the first shell of $\vec{G}$ reciprocal lattice vectors with their corresponding
numeral labels used in the main text. The three circles in red correspond to the $\vec{G}_0, \vec{G}_{\pm}$ vectors with 
large inter-layer tunneling coefficients. The three vectors $\vec{K}_j = \vec{K} + \vec{G}_j$ $j=0, \pm$ 
have the same magnitude.
{\em Right Panel:} First shell moir{\' e} reciprocal lattice vectors $\widetilde{G} = -\theta \hat{z}\times \vec{G}$
for graphene on graphene result differ from the honeycomb reciprocal lattice 
vectors by a clockwise 90$^{\circ}$ rotation and a reduction in size by a factor proportional to $\theta$.
The solid black arrows represent the $\vec{Q}_j$ vectors
that connect a $\vec{q}$ vector in the bottom layer to one in the top layer
and the red hexagon encloses the moir{\' e} pattern Brillouin-zone.  
}
\label{moirecell0}
\end{figure}

Our low energy model is naturally employed in combination with a 
continuum model in which wave vectors are measured from the Dirac point.  
The condition that $\vec{k'} = \vec{k} + \tilde{ G}$ then translates into the condition   
\begin{equation} 
\vec{q'}=\vec{q} + \vec{K}-\vec{K'} + \tilde{ G} = \vec{q} + \vec{Q}_j    
\end{equation}
where $j=0,\pm$ and the indices correspond to the three $\vec{G}$'s for which 
$H_{ls,l's'}(\vec{K}:\vec{G}$) is large: $\vec{G}=(0,0)$ and  
\begin{equation}
G_{\pm} = \frac{4\pi}{\sqrt{3}a} \, ( - \frac{\sqrt{3}}{2}, \pm \frac{1}{2}) = K (- \frac{3}{2}, \pm \frac{\sqrt{3}}{2}).
\end{equation} 
Here $K$ is the magnitude of the Dirac wave vector.  Taking account of the difference between the rotated and 
unrotated system reciprocal lattices we find that to first order in $\epsilon$ and $\theta$ 
\begin{equation} 
\label{qvectors}
\vec{Q}_j =  \epsilon \vec{K}_j - \theta \hat{z} \times \vec{K}_j
\end{equation} 
where $\vec{K}_j = \vec{K} + \vec{G}_j$.  Note that, independent of the values of 
$\theta$ and $\epsilon$, the three vectors $\vec{Q}_{j}$ have the same
magnitude $K \sqrt{\epsilon^2 + \theta^2} $
and that they are related by $120^{\circ}$ rotations.    
In the graphene case the parameter $\epsilon$ that accounts 
for the difference in lattice constant between the layers is equal 
to zero, but we retain it here because of the close similarity between the G/G 
interlayer hopping terms and the G/BN cases discussed below.  
When momenta are measured from the Dirac point, a state in one-layer is coupled 
to states in the same layer separated in momentum space by moir{\' e} pattern reciprocal lattice 
vectors, and to states in the opposite layers separated by 
moir{\' e} pattern reciprocal lattice vectors $\pm \vec{Q}_{j}$. 
(See Fig.~\ref{moirecell0}.)  

For G/G the intra-layer contribution to the Hamiltonian is negligible for $\vec{G} \ne (0,0)$, and 
for $\vec{G} = (0,0)$ and $\vec{k}=\vec{K}$ its dependence on site labels is proportional to a unit matrix.  
It follows that the $\vec{G} = (0,0)$, $\vec{k}=\vec{K}$ Hamiltonian can be set to zero 
by choosing the zero of energy appropriately.   
The dependence of the $\vec{G} = (0,0)$ interlayer Hamiltonian on $\vec{k}$ satisfies the same 
symmetry requirements as the isolated layer Hamiltonian.  
We have found that for both G/G and G/BN cases, the 
difference between the $\vec{G} = (0,0)$ interlayer Hamiltonian and the isolated layer 
Hamiltonian is negligible.  

\begin{figure}[htbp]
\label{ggaa}
\includegraphics[width=8cm,angle=0]{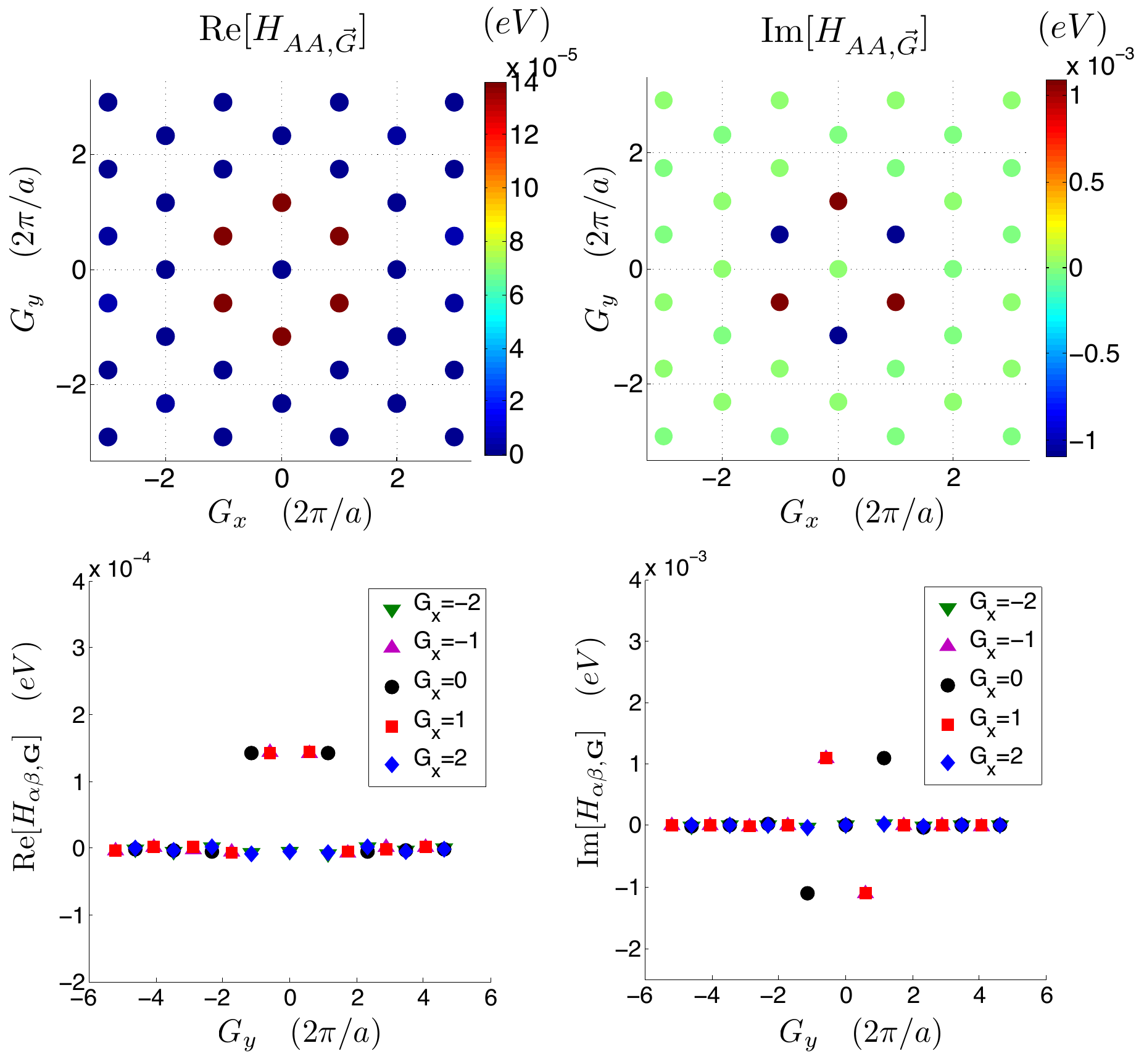}  
\caption{
(Color online)
Real and imaginary parts of the Fourier transforms of the 
intra-layer AA site diagonal Hamiltonian matrix element.  Because of the symmetries of the honeycomb lattice 
the $\vec{G}$-dependent site potentials satisfy 
$H_{AA, {\vec{G}}} = H^*_{BB, {\vec{G}}} =  H^*_{A'A', {\vec{G}}} = H_{B'B', {\vec{G}}}$. 
These contributions to the twisted layer Hamiltonian for G/G are small and can
often be neglected.  
}
\end{figure}

When only the largest non-zero 
interlayer coupling terms are retained,
the Hamiltonian is the sum of three terms, each a product of a coupling constant $t_{bt} = 113 {\rm meV}$, 
 a momentum boost factor $\delta_{\vec{q}',\vec{q}+\vec{Q}_j}$ 
and  a sublattice dependent factor
\begin{equation} 
T^{j}_{s,s'} =   t_{bt} \exp(i\vec{G_j} \cdot (\vec{\tau}_{s} - \vec{\tau}_{s'}  - \vec{\tau})), 
\end{equation} 
where we have restored the phase change due to the translation $\vec{\tau}$ prior to rotation.
When the momentum boost operator is written in real space it it local and has a plane-wave 
spatial dependence.  We therefore obtain a Hamiltonian with a space-dependent inter-layer coupling 
Hamiltonian that has a sub lattice pseudo-spin dependence:
\begin{equation}
{ H}_{bt}(\vec{r}) =    \, \sum_{j} \,  \exp(-i \vec{Q}_j \cdot \vec{r} ) \, T^{j}_{s,s'}
\end{equation} 
where 
\begin{equation}
 T^{j}=   t_{bt} \exp(- i \vec{G}_j  \vec{\tau})
\left(\begin{matrix}1 & \exp(-{i j\phi}) \\ \exp(i j\phi) & 1\end{matrix}\right)
\label{tmatrices}
\end{equation}
and $\phi=2\pi/3$.  
A similar formula was derived previously starting from {\em ad hoc} $\pi$-band tight-binding models 
\cite{Santos07,RafiMoireBand}.   
(Note that in Ref. [\onlinecite{RafiMoireBand}] the initial displacement $\tau$ 
was defined relative to $AB$ stacking.) 
This position-dependent inter-layer tunneling can be understood in terms of  
local interlayer coordination which varies with the moir{\' e} periodicity between $AA$, $AB$ and 
intermediate arrangements. 
Here we demonstrate by explicit first-principles calculations that this model
for twisted layer electronic structure is quite accurate.
Our {\em ab initio} calculations give rise to a coupling constant of 
$t_{bt} = 113 {\rm meV}$, nearly identical to the value
$t_{bt} = 110$meV estimated previously by fitting tight-binding models to 
the experimentally known Dirac point spectrum of bilayer graphene.
In the phenomenological tight-binding model context, the applicability of 
this model was justified on the basis of the argument\cite{RafiTun} 
that any reasonable inter-layer tunneling {\em ansatz} yields a dependence on two-dimensional
position that is smooth at atomic scale.  Our microscopic calculations free us 
from an {\em ad hoc} tight-binding model and confirm the expected smoothness.
The success of the {\em ad hoc} tight-binding model in describing 
interlayer tunneling effects may be traced to the property that 
only one number, namely $t_{bt}$, is important for the low-energy 
electronic structure. Any microscopic model which is adjusted so that it gives 
an appropriate value for $t_{bt}$ will yield similar predictions.    

The  explicit form of the $\vec{d}$-dependent inter-layer Hamiltonian which retains only the 
single strong moir{\' e} band model parameters follows from Eq.~(\ref{Hofd}):
\begin{eqnarray}
\label{gginterlayer}
H_{AA'}(\vec{K} :\vec{d}) &=& H_{BB'}(\vec{K} :\vec{d}) =  \\
&=& t_{bt} \left( 1 + \exp(-i \vec{G}_+ \cdot \vec{d}) + \exp(-i \vec{G}_- \cdot \vec{d})  \right)    \nonumber   \\
H_{AB'}(\vec{K} :\vec{d}) &=&    
t_{bt} \left( 1 + \exp(-i \phi  )  \exp(-i \vec{G}_+ \cdot \vec{d})  \right.     \nonumber    \\
&+&  \left. \exp(i \phi )  \exp(-i \vec{G}_- \cdot \vec{d})  \right)        \nonumber  \\
H_{BA'}(\vec{K} :\vec{d}) &=& 
t_{bt} \left( 1 + \exp(i \phi )  \exp(-i \vec{G}_+ \cdot \vec{d})  \right.       \nonumber  \\
&+&   \left. \exp(-i \phi )  \exp(-i \vec{G}_-  \cdot \vec{d})  \right)         \nonumber 
\end{eqnarray}
Although they are relatively weak in the G/G case,  for completeness we specify the leading  
intralayer terms as well.  
The Fourier transforms of the intra-layer matrix elements 
have the same magnitude within a shell of reciprocal lattice vectors.  
Including the first shell only, we obtain the following parametrization of the 
$\vec{d}$-dependent intra-layer Hamiltonian matrix elements:  
\begin{eqnarray}
\label{gbn_intra}
H_{ii}(\vec{K} :\vec{d}) &=&  C_{0 ii} + 2 C_{ii} \, {\rm Re}[ f(\vec{d})  \exp[i \varphi_{ii}]  ] ,  \\
H_{ij}(\vec{K} :\vec{d}) &=& { g(C_{ij}, \varphi_{ij}) }, \quad \quad  (  i \neq j )     \nonumber
\end{eqnarray}
where 
\begin{eqnarray}
f (\vec{d}) = \exp[- i G_1 d_y]  + 2 \exp[i \frac{G_1 d_y}{ 2}] \cos(\frac{\sqrt{3}}{2}G_1 d_x),
\end{eqnarray}
$G_1 = 4 \pi/ \sqrt{3}a $.
The matrix elements labelled by $ii = AA, BB, A'A', B'B'$, are the $\vec{d}$-dependent 
site energies.  The matrix elements labelled by $AB$ and $A'B'$ describe inter-sublattice 
tunneling at the Dirac point within the layers.  In Eq.~(\ref{gbn_intra})    
\begin{eqnarray}
g(C, \varphi) &=& 2 C  \cos \left(\frac{\sqrt{3} G_1}{2}d_x   \right)
\cos \left(\frac{G_1}{2}d_y - \varphi  \right) \quad        \\ 
&-&   2C \cos \left( G_1 d_y + \varphi \right)    \nonumber \\
&-& i 2\sqrt{3} C \sin \left( \frac{ \sqrt{3}G_1}{2} d_x  \right) \sin \left( \frac{ G_1}{2} d_y  - \varphi \right) .  \nonumber 
\end{eqnarray}
Using the numerical labels for the $\vec{G}$ vectors in Fig. \ref{moirecell0}
\begin{eqnarray}
\label{relationcoefdiag}
H_{ii, {\vec{G}_1}} &=& H_{ii, {\vec{G}_3}} = H_{ii, {\vec{G}_5}} = C_{ii} \exp(i \varphi_{ii}), \\
H_{ii, {\vec{G}_2}} &=& H_{ii, {\vec{G}_4}} = H_{ii, {\vec{G}_6}} = C_{ii} \exp(- i \varphi_{ii})  \nonumber
\end{eqnarray}
for the diagonal terms, and
\begin{eqnarray}
\label{relationcoefoffdiag}
H_{A^{(\prime)}B^{(\prime)}, \vec{G}_1} &=& H^{*}_{A^{(\prime)}B^{(\prime)}, \vec{G}_4} = C_{A^{(\prime)}B^{(\prime)}} \exp(i (- \varphi_{AB} - \pi )),   \\
H_{A^{(\prime)}B^{(\prime)}, \vec{G}_3} &=& H^{*}_{A^{(\prime)}B^{(\prime)}, \vec{G}_2} = C_{A^{(\prime)}B^{(\prime)}} \exp(i (  -\varphi_{AB} + \pi/3)),   \nonumber \\
H_{A^{(\prime)}B^{(\prime)}, \vec{G}_5} &=& H^{*}_{A^{(\prime)}B^{(\prime)}, \vec{G}_6} = C_{A^{(\prime)}B^{(\prime)}} \exp(i (  -\varphi_{AB}  - \pi/3 ))  \nonumber
\end{eqnarray}
for the off diagonal terms.
For graphene on graphene, the expansion coefficients satisfy
the symmetry properties 
$H_{AA, {\vec{G}}} = H^*_{BB, {\vec{G}}} =  H^*_{A'A', {\vec{G}}} = H_{B'B', {\vec{G}}}$ for the diagonal terms
and $H_{AB, {\vec{G}}} = H_{A'B', {\vec{G}}}$ for the off diagonal terms, $C_{0AA} = C_{0BB}$,
and  $\varphi_{AB} = \varphi_{A'B'} = 0$.
The numerical values of the nonzero parameters that define the model for G/G are  
\begin{eqnarray}
\label{paramsgg}
C_{AA} &=& C_{B'B'} = 1.10 \,\, {\rm meV}, \,\, \,\,\, \varphi_{AA} = \varphi_{B'B'} = 82.54^{\circ},   \quad \\
C_{BB} &=& C_{A'A'} = C_{AA} \,\, , \quad \,\,\, \varphi_{BB} = \varphi_{A'A'} = - \varphi_{AA},    \nonumber \\
C_{AB} &=& 2.235 \,\, {\rm meV}\, .     \nonumber 
\end{eqnarray}
The sublattice site-energy difference $2 H_z ( \vec{K}: \vec{d}) =  H_{AA}(\vec{K}: \vec{d}) -   H_{BB}(\vec{K}: \vec{d}) $
vanishes for AA stacking and reaches its maximum value $\sim$12 meV for the AB stacking configuration.
This value is in reasonable agreement with the $\sim$15 meV site-energy difference estimated
elsewhere for AB stacked bilayer graphene.\cite{bilayertb} 
It will be interesting to see if these relatively small terms which are 
normally neglected in two-layer graphene systems
have any observable consequences.  
Eq. (\ref{paramsgg}) also implies spatial variations of the average site-energy
$ H_0 ( \vec{K}: \vec{d}) =  ( H_{AA}(\vec{K}: \vec{d}) +   H_{BB}(\vec{K}: \vec{d}) )/2$ 
that are smaller than 1 meV.  These variations will tend to drive
small charge transfers between different parts of the moir{\' e} pattern, but 
their role is not especially important because of their small value. 

\begin{figure}[htbp]
\label{ggab}
\begin{tabular}{c}
\includegraphics[width=8cm,angle=0]{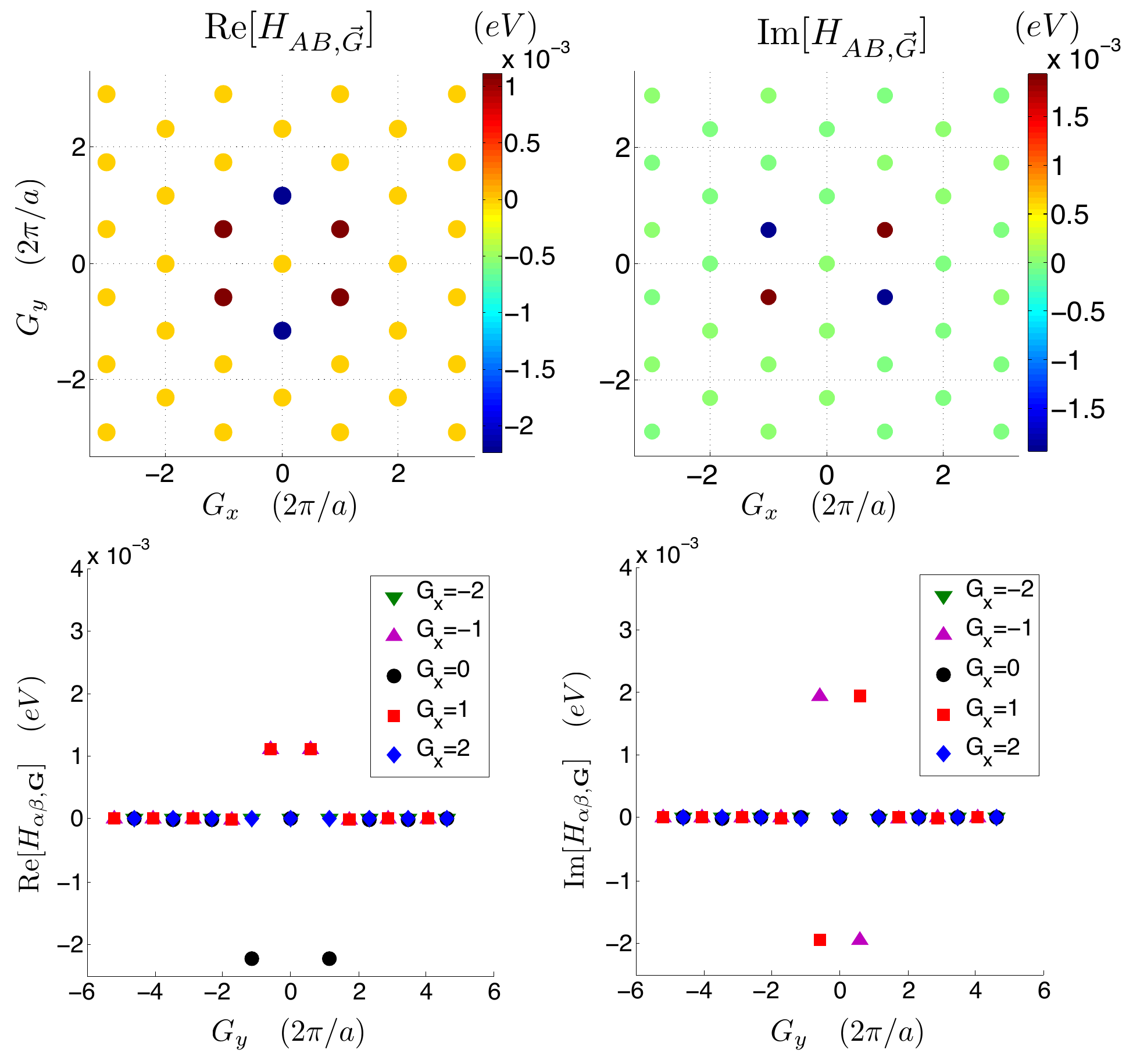} 
\end{tabular}
\caption{
(Color online)
Real and imaginary parts of the Fourier transform 
of the intralayer, intersubband Hamiltonian matrix element $H_{AB,\vec{G}}$.
It follows from symmetry that $H_{AB,\vec{G}} = H_{A'B',\vec{G}}$.
These matrix elements are also small. 
}
\end{figure}

\subsubsection{First shell approximation for commensurate AB, AA limits}

\begin{figure} 
\begin{tabular}{c}
\includegraphics[width=8.4cm,angle=0]{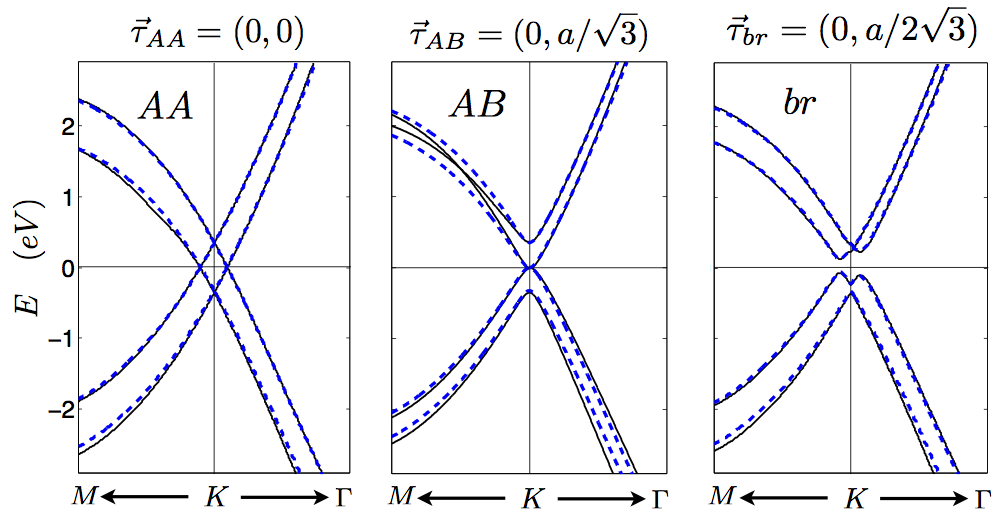} 
\end{tabular}
\caption{
(Color online) 
Comparison of LDA G/G band structures (solid lines) and
the one-parameter moir{\' e} band model (blue dashed lines), 
which retains only the first shell Fourier-expansion of inter-layer coupling. 
Results are shown for commensurate G/G with  
AA, AB, and $\vec{\tau}_{br}=(0, a/2\sqrt{3})$ bridge stacking. 
The electronic structure for BA stacking is identical to AB stacking.
We find excellent agreement between the direct and approximate calculations,
demonstrating the accuracy of the first 
shell approximation for interlayer coupling.
The intralayer tight-binding Hamiltonian uses the models in Refs. [\onlinecite{bilayertb,graphenetb}] with the experimental lattice constants
of $a = 2.46~\AA$ whereas the interlayer coupling is given by the first shell approximation as parametrized in Eq.~(\ref{gginterlayer}).
The small differences can be attributed mainly to the approximations involved in the first shell approximation for describing the
interlayer coupling.
}
\label{ggcommensurate}
\end{figure}

In the following we test the single-parameter moir{\' e} band model 
in which the inter-layer Hamiltonian is truncated at the first shell of its Fourier 
expansion by  applying it to the crystalline AA and AB stacking limits. 
In the crystalline limit, $\vec{d}$ is independent of position
and $\vec{Q}_j=(0,0)$ for $j=0,\pm$. 
In the AA stacking configuration, interlayer coupling is maximized because  
carbon atoms in different layers sit exactly on top of each other.
Mirror symmetry leads to layer-symmetric and layer-antisymmetric 
copies of the single-layer Dirac spectrum.
With our conventions, AA stacking corresponds to $\vec{d}=(0,0)$ independent of $\vec{L}$, 
and to  
Direct evaluation of the AA Wannier matrix elements yields 
$H_{AA^{\prime}}(\vec{K} : \vec{d}=(0,0)) = 
H_{BB^{\prime}}(\vec{K} : \vec{d}=(0,0)) = 355 $ meV and 
$H_{AB^{\prime}}(\vec{K} : \vec{ d}=(0,0)) = H_{BA^{\prime}}(\vec{K} : \vec{d}=(0,0)) =  0$.  
The matrix element from the Fourier expansion model truncated at the first shell is $ 3 t_{bt} = 339$ meV.   
For the case of AB stacking ($\vec{d} = (0, a/\sqrt{3})$, a direct evaluation of the
Wannier matrix elements yields 
$ H_{BA^{\prime}}(\vec{K} : \vec{d} = (0, a/\sqrt{3})) = 354$ meV,
with all other interlayer coupling elements vanishing. 
(The small difference of 7 meV with respect to the calculation for AB bilayer graphene presented in Ref. [\onlinecite{bilayertb}] 
is due to the slightly smaller in-plane lattice constants used here.)
These comparisons demonstrate that the truncated Fourier expansion 
single-parameter model yields matrix elements that are 
are typically inaccurate by $\sim 15$ meV, or by around 5\% in relative terms. 
We emphasize that the truncation at the first shell in the Fourier expansions 
is not essential to our approach, but is attractive because it yields a 
model that is specified by a single parameter. 
The approximate band structures obtained using the above interlayer coupling matrices 
are compared against the first principles LDA bands in Fig. \ref{ggcommensurate}.

\begin{figure}
\begin{center}
\includegraphics[width=8.5cm]{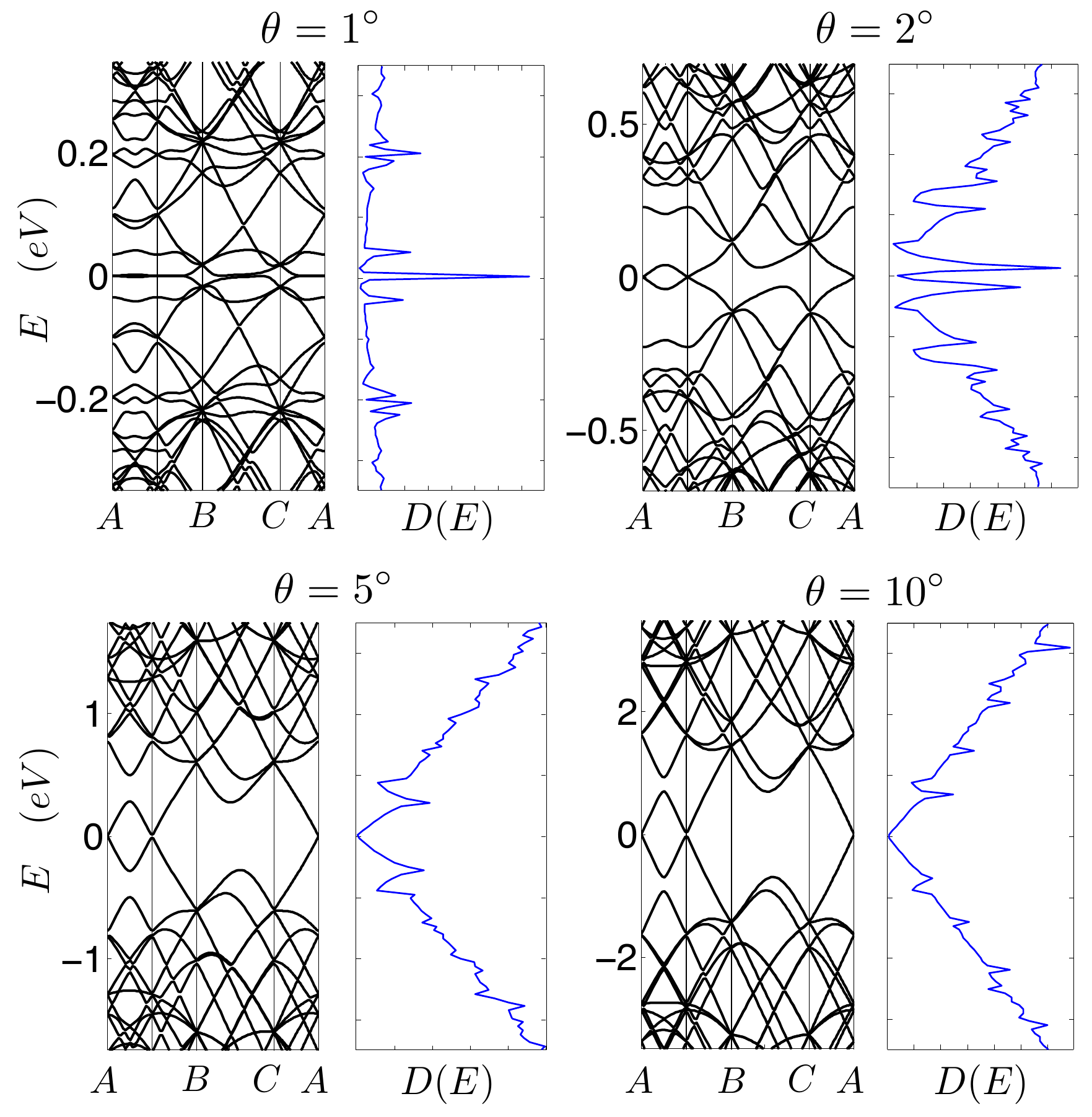} 
\end{center}
\caption{  Moir{\' e} band structure and density of states of two graphene layers for four different 
relative orientation angles.  Our results are similar to those obtained in 
Refs. [\onlinecite{RafiMoireBand,SantosII}].
We plot the band structure as a function of 
momentum along the straight lines in $k$-space connecting points A, B, C and A
in Fig. \ref{moirecell0}. The accompanying density-of-states plots demonstrate the 
complex influence of interlayer coupling, which is responsible for many 
van Hove singularities.
}
\label{ggmoirebands}
\end{figure}

\subsubsection{Application to twisted bilayer graphene}
For G/G we have $\varepsilon = 0$ 
so that the moir{\' e} pattern reciprocal lattice vectors are related to the 
honeycomb reciprocal lattice vectors by 
\begin{eqnarray}
\widetilde{G} = -\theta \,  \hat{z} \times \vec{G}.
\end{eqnarray}
The Hamiltonian matrix for a given wave vector $\vec{k}$ in
the moir{\' e} Brillouin-zone (MBZ) can be constructed using
Eq.~(\ref{matrixelementbb}).  
The momentum boost operators in the inter-layer Hamiltonian terms 
connect states whose momenta differ by $\vec{Q}_j$,
while those in the intra-layer Hamiltonian terms connect states
whose momenta differ by $\widetilde{G}$.
From Eq.~(\ref{qvectors}) the explicit expression for the $\vec{Q}_j$'s is 
\begin{eqnarray}
\vec{Q}_{0} &=& \theta K (0, -1)  \label{q1} \\
\vec{Q}_{+} &=& \theta K ( -\frac{\sqrt{3}}{2},   \frac{1}{2})   \label{q2}    \nonumber \\
\vec{Q}_{-}  &=& \theta K ( \frac{\sqrt{3}}{2} ,  \frac{1}{2})   \label{q3} .    \nonumber
\end{eqnarray}

For every $\vec{k}$ in the MBZ we can construct matrices  
with $2 \times 2$ sub-lattice blocks.  The isolated layer Dirac 
Hamiltonian contributes blocks that are diagonal in wave vector and layer. 
The blocks that account for tunneling 
from bottom to top layers involve momentum boosts by  
$\vec{Q}_{j}$ whereas those that connect the same layer involve
momentum boosts by a moir{\' e} pattern reciprocal lattice vector.  
Since the $\vec{Q}_{j}$'s change sign with tunneling direction, and the 
difference between any pair of $\vec{Q}_{j}$'s is a moir{\' e} pattern reciprocal lattice vector 
(see Fig.~\ref{moirecell0}),
the crystal momentum defined by the  moir{\' e} pattern periodicity is a good quantum number.
For every $\vec{k}$ in the moir{\' e} pattern Brillouin-zone, a finite 
matrix can be constructed by cutting off the plane-wave expansion.

The moir{\' e} bands obtained by 
diagonalizing the matrix constructed in this way are plotted in Fig.~\ref{ggmoirebands}.
We find bands that are similar to those described in 
Refs.~(\onlinecite{RafiMoireBand,SantosII}) in which moir{\' e} bands
were derived from phenomenological tight-binding models
rather than form {\em ab initio} DFT calculations.   The close agreement 
is expected since both models are accurately approximated by a model
with a single interlayer tunneling parameter, as explained above.  
We now turn to the G/BN case in which the layer coupling effects are 
more complex.  There we will see that our approach, 
which provides a route to build an effective model based on 
DFT bands, has distinct advantages over a purely phenomenological approach.

\subsection{Graphene on Boron Nitride} 
\label{sec:grapheneonbn} 

\subsubsection{Moir{\' e} Band Model} 

\begin{figure} 
\begin{tabular}{c}
\includegraphics[width=8.4cm,angle=0]{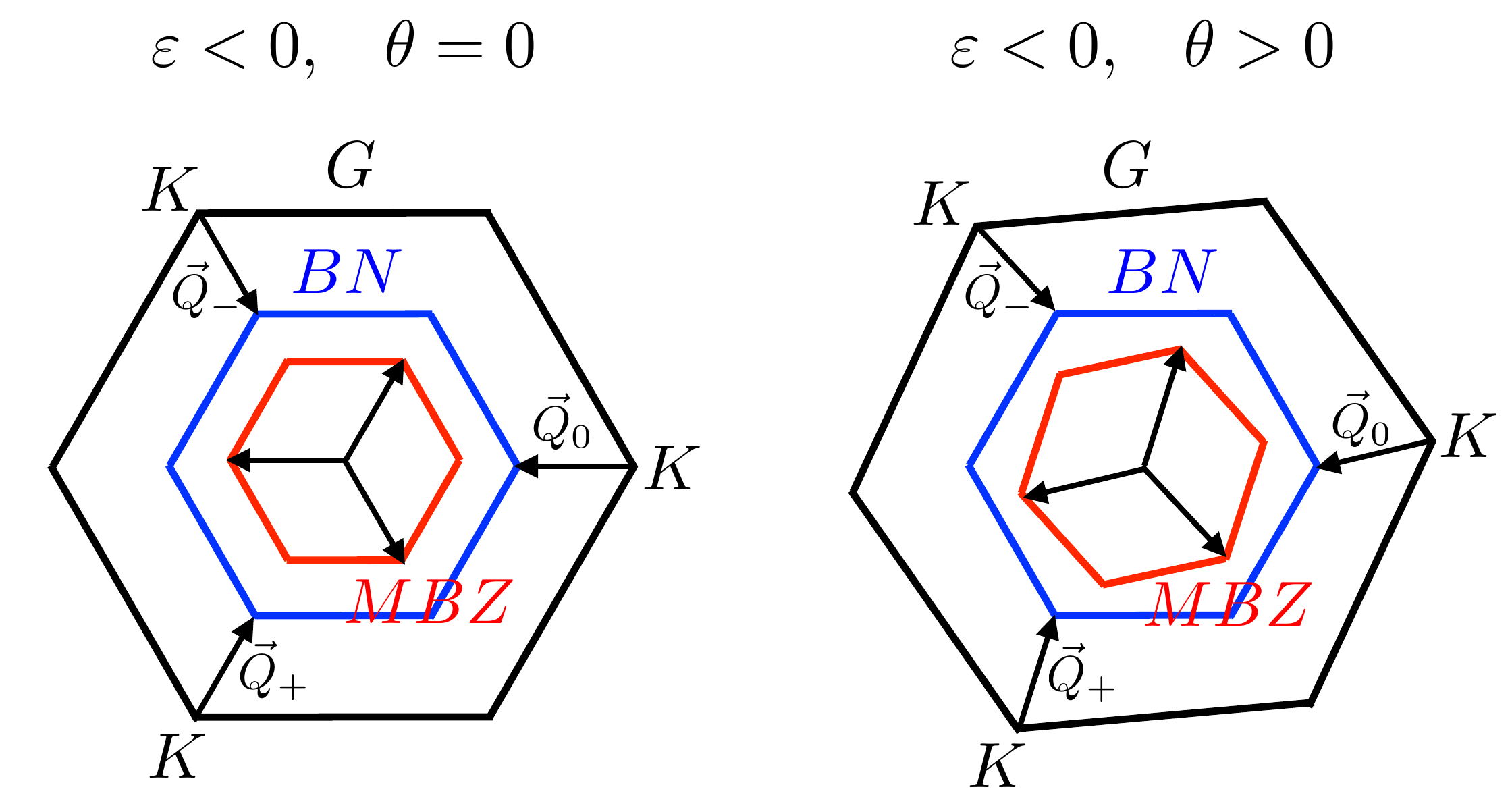} 
\end{tabular}
\caption{
(Color online)
Schematic Brillouin zones of graphene in black lines, of hBN in blue, and the moir\'e Brillouin zone (MBZ) in red
for incommensurate G/BN. The difference between the lattice constants has been exaggerated to aid visualization.
Both the size and orientation of the MBZ change continuously with twist angle
due to the lattice constant mismatch. 
The three $\vec{Q}_j$ vectors given in Eq.~(\ref{qvectors}) connect the $K$-points
of graphene to those of hBN. 
The moir{\' e} pattern reciprocal lattice vectors can be constructed by 
summing pairs of $\vec{Q}_j$ vectors.}
\label{hbnbzone}
\end{figure}

The crystalline lattices we used to derive the moir{\' e} band model parameters for G/BN  
were identical to those used for the G/G case, except that the bottom layer was changed from graphene to hBN.
Because hBN has a slightly larger lattice constant than graphene, 
the moir\'e pattern reciprocal lattice vectors are in this case given by the 
more general expression (Eq.~(\ref{tildeG})) which accounts for both dilation and twist. 
The moir\'e pattern Brillouin zone therefore continuously changes its orientation 
as a function of twist angle $\theta$ as we illustrate in Fig. \ref{hbnbzone}.

\begin{figure*} 
\begin{center}
\includegraphics[width=18cm,angle=0]{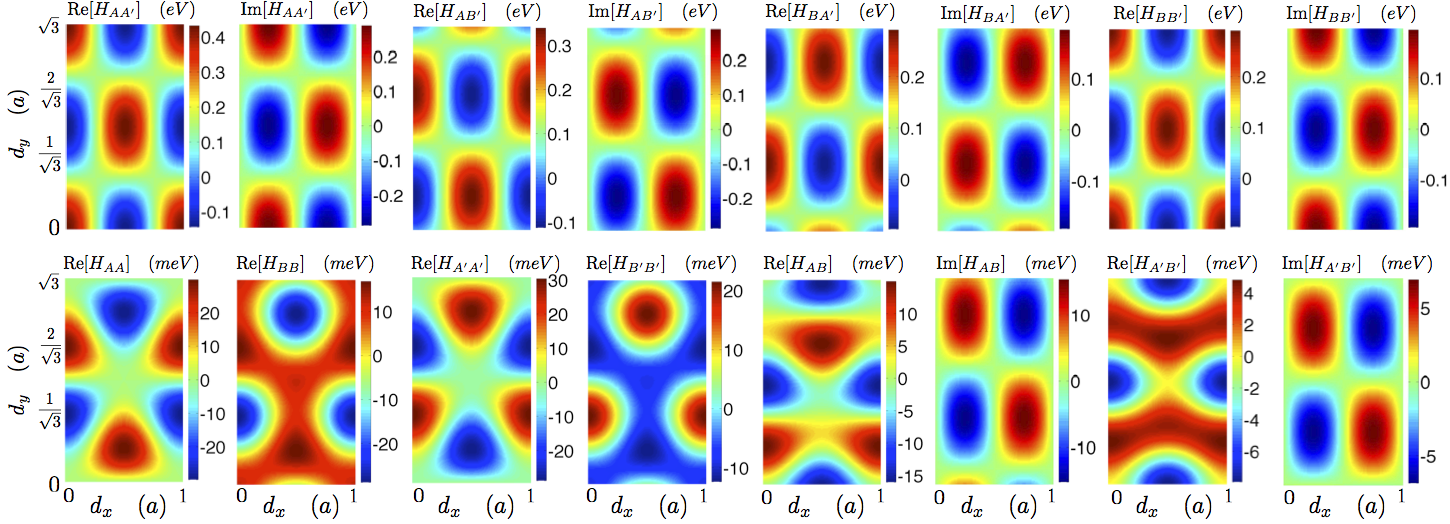}
\end{center}
\caption{
(Color online)
Displacement vector $\vec{d}$ dependence of 
Wannier representation interlayer 
Hamiltonian matrix elements for G/BN.
{\em Upper panel:}
Matrix elements $AA'$, $AB'$ connecting boron 
with carbon, and matrix elements $BA'$, $BB'$ connecting nitrogen
with carbon. The interlayer coupling 
matrix elements vary over a large range $\sim 600$ meV.
{\em Lower panel:}
Displacement  vector $\vec{d}$ dependence of 
intra-layer Wannier representation 
Hamiltonian matrix elements for G/BN.
On-site energies in the graphene layers vary by $\sim$ 60 meV.
In these plots the site energies are plotted relative to their 
spatial averages $C_{0ii}$. (See Eq.~(\ref{gbnhamiltonian}).)
The carbon layer AB inter-sublattice terms vary over a range of $\sim$ 35 meV
whereas the BN layer  $A'B'$ terms vary by $\sim$ 15 meV. 
}
\label{interlayer2}
\end{figure*}

\begin{figure}[htbp]
\begin{center}
\includegraphics[width=8.5cm,angle=0]{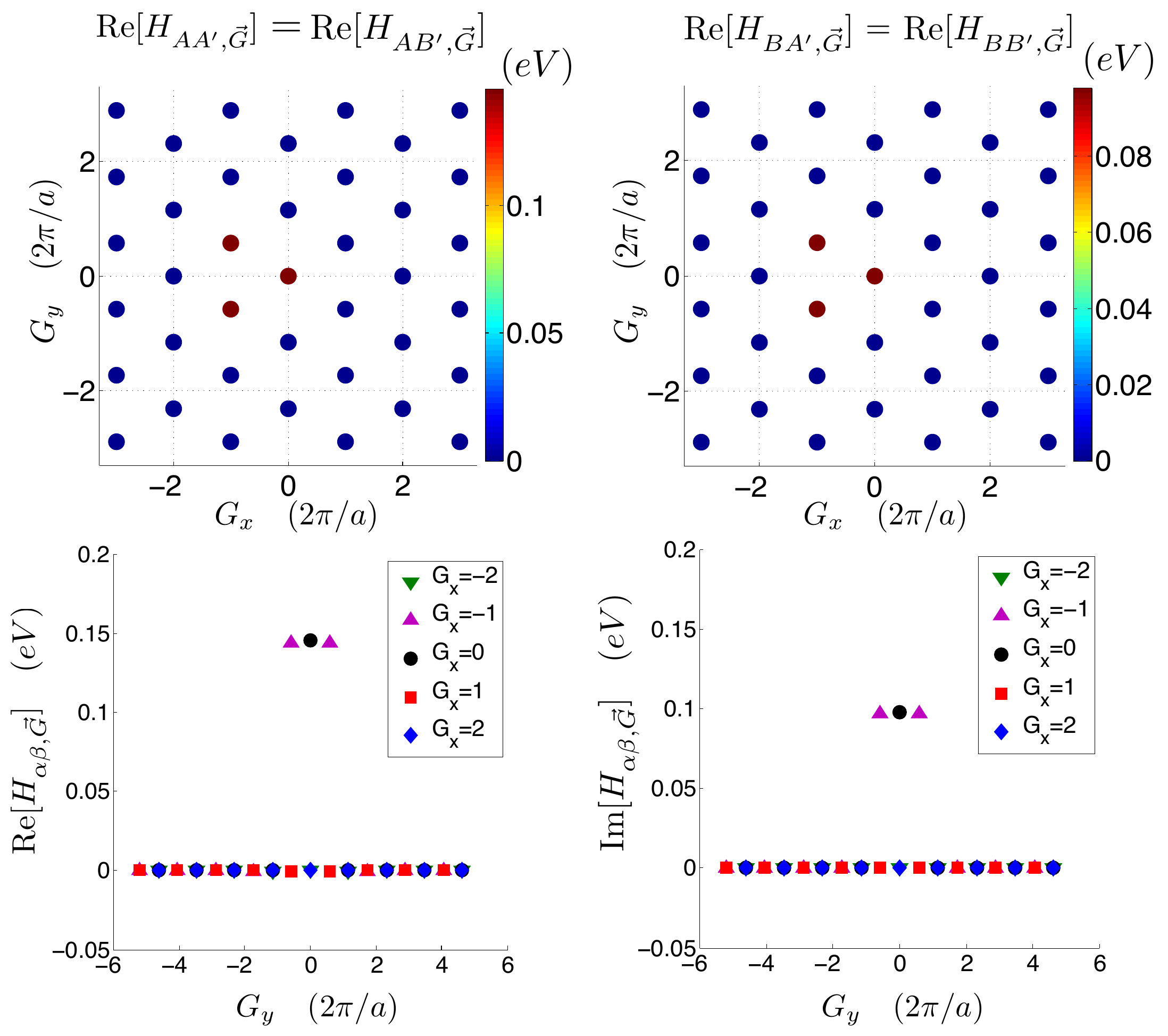} 
\end{center}
\caption{(Color online)
G/BN interlayer moir{\' e} band model parameters 
obtained by evaluating Fourier expansion coefficients for the 
layer separation dependence of Dirac point Wannier representation Hamiltonian 
matrix elements.  As in the G/G case three Fourier coefficients 
dominate interlayer coupling.    
}
\label{gbninterlayer}
\end{figure}

\begin{figure*}[htbp]
\begin{center}
\begin{tabular}{cc}
\includegraphics[width=8.5cm,angle=0]{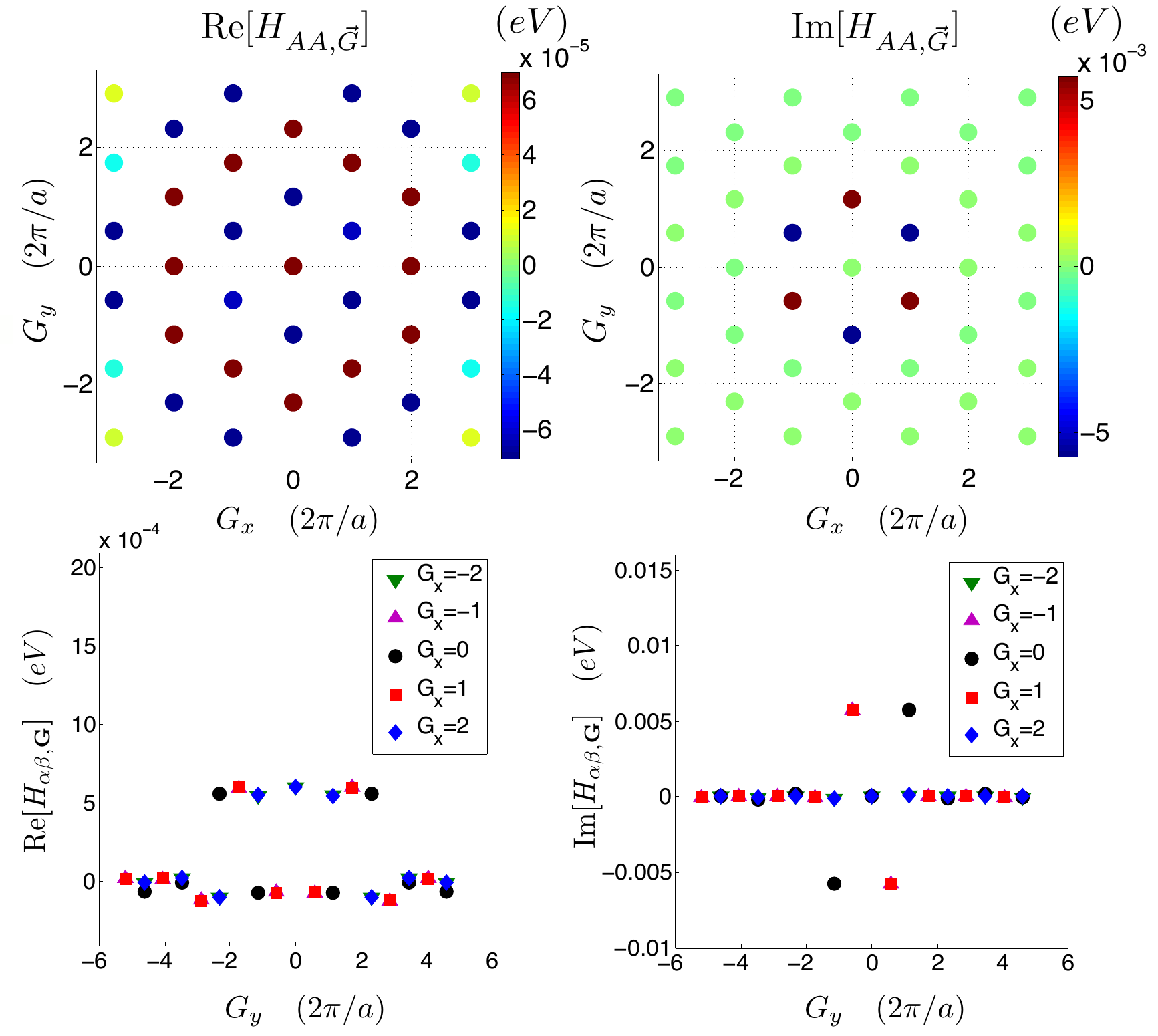} \quad & \quad
\includegraphics[width=8.5cm,angle=0]{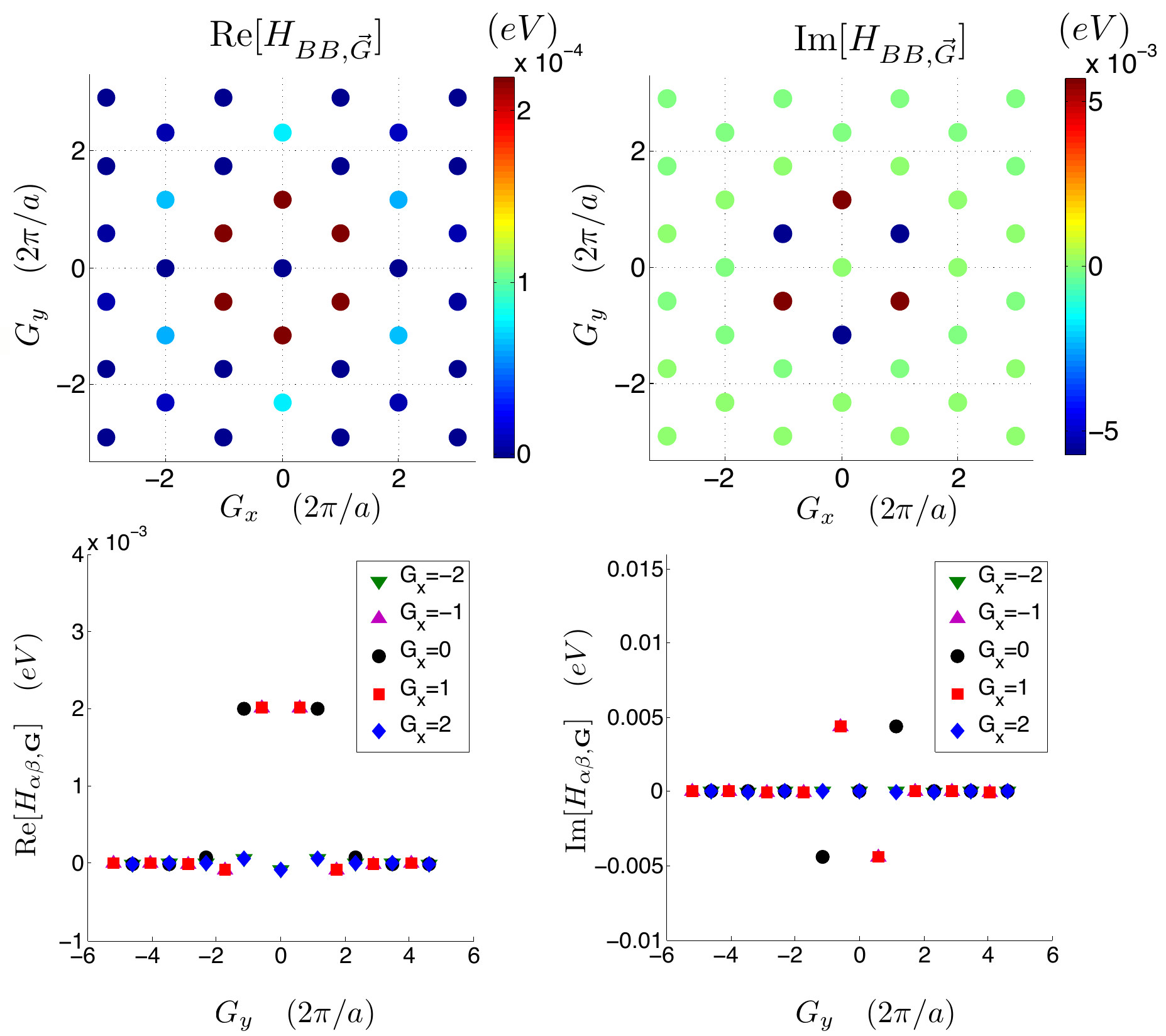} \\ & \\
\includegraphics[width=8.5cm,angle=0]{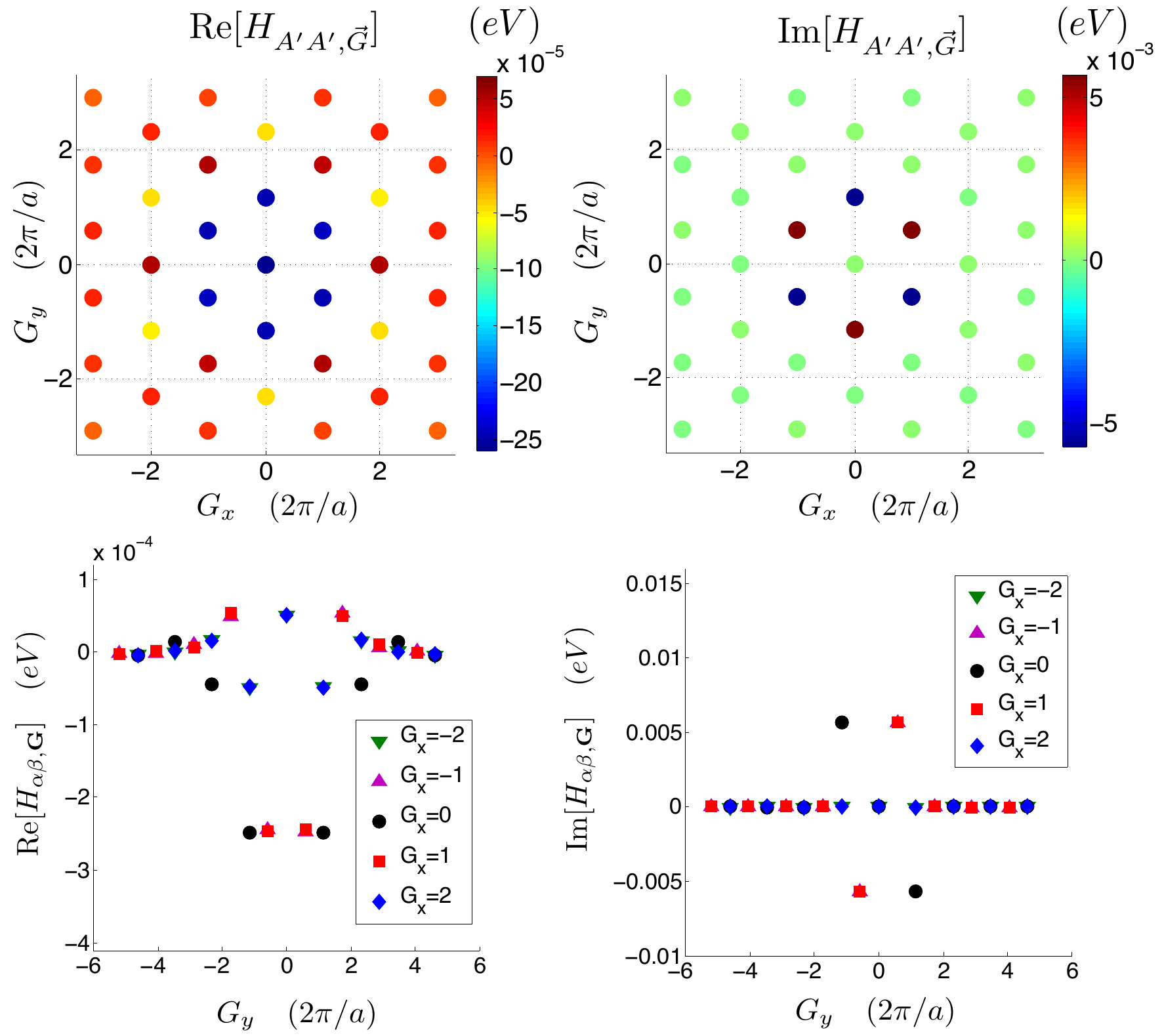}  \quad & \quad
\includegraphics[width=8.5cm,angle=0]{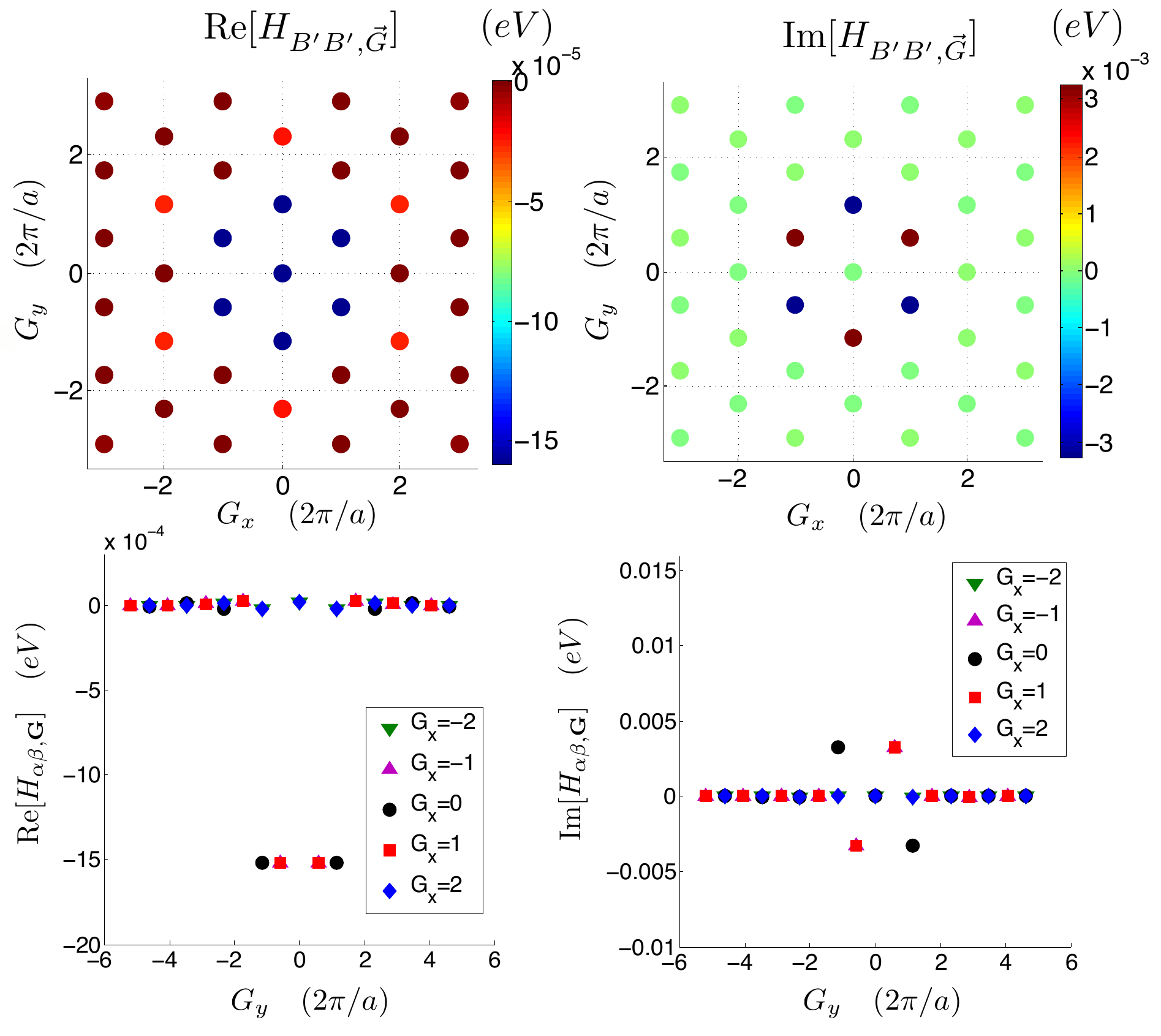} \\
\end{tabular}
\end{center}
\caption{
Fourier expansion of the Wannier-representation matrix-elements
$  H_{AA} (\vec{K} : \vec{d}) $,
$  H_{BB} (\vec{K} : \vec{d}) $,
$  H_{A'A'} (\vec{K} : \vec{d}) $ and
$  H_{B'B'} (\vec{K} : \vec{d}) $ that describe the interlayer displacement
dependence of site-energies in the hBN and graphene layers.
These numerical results demonstrate that the site energies are 
accurately approximated by the model that includes only the first shell of reciprocal 
lattice vectors.  The parameters of this model are listed in the main text.   
We have chosen the average energy on the carbon sites as the zero of energy.
With this choice, the elements corresponding to $\vec{G} = 0$ are
$H_{AA} (\vec{K} : \vec{G} = 0 ) = 3.332$ eV for boron,
$H_{BB} (\vec{K} : \vec{G} = 0 ) = -1.493$ eV for nitrogen,
$H_{A'A'} (\vec{K} : \vec{G} = 0 ) = 0$ eV 
and $H_{A'A'} (\vec{K} : \vec{G} = 0 ) = 0$ eV. 
}
\label{gbnintrasite}
\end{figure*}
\begin{figure*}[htbp]
\begin{center}
\begin{tabular}{cc}
\includegraphics[width=8.5cm,angle=0]{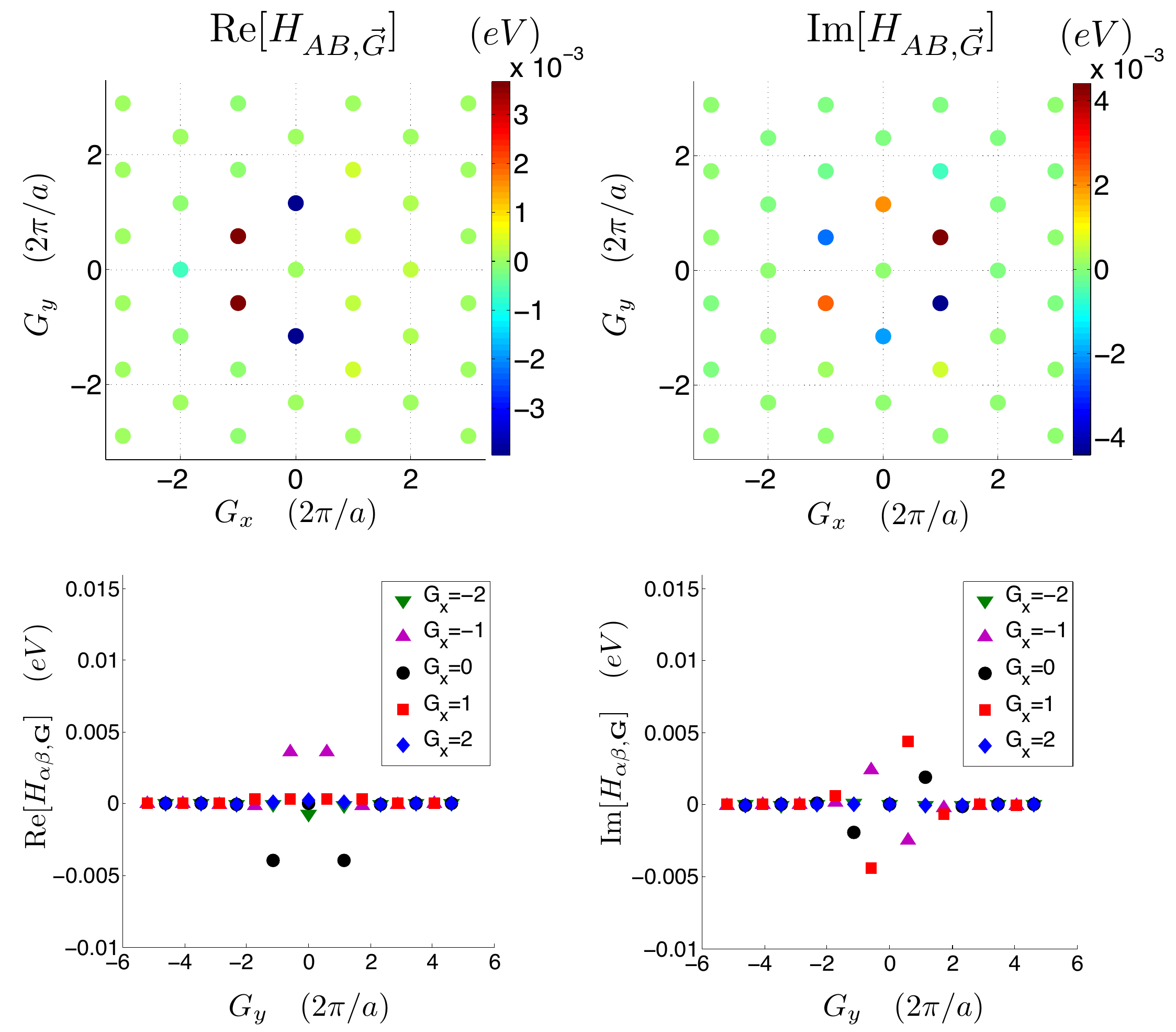} \quad & \quad
\includegraphics[width=8.5cm,angle=0]{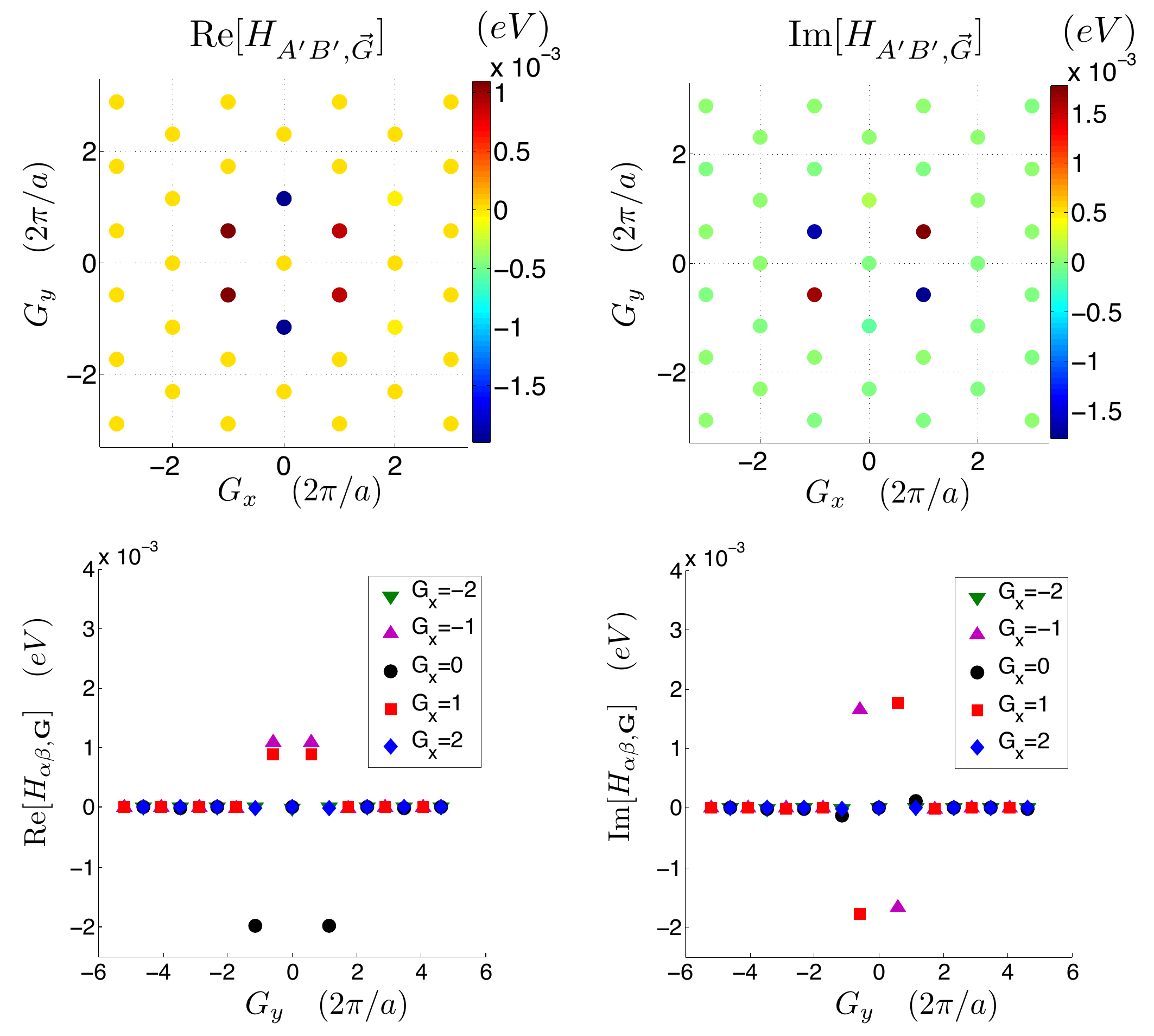} 
\end{tabular}
\end{center}
\caption{(Color online)
Fourier expansion of the Wannier-representation matrix-element
$  H_{AB} (\vec{K} : \vec{d}) $, which describes inter-sublattice tunneling within the hBN layer, and
$  H_{A'B'} (\vec{K} : \vec{d}) $, which describes inter-sublattice  tunneling within the graphene layer.
These numerical results demonstrate that the local coordination dependence of 
interlayer hopping processes is accurately approximated by the model that includes only the first shell of reciprocal 
lattice vectors.  The parameters of this model are listed in the main text.   
}
\label{gbnintralayer}
\end{figure*}

In order to capture the local coordination dependence of the 
electronic structure we have evaluated Wannier-representation 
bands over the complete range of inter-layer displacement $\vec{d}$ values.
The dependence on $\vec{d}$ of Dirac point matrix elements 
is summarized in Fig. \ref{interlayer2}.  As in the G/G case, these {\em ab initio} 
results provide the 
chemical information that we use to construct a moir{\' e}-band model that 
can account for the lattice constant difference between graphene and hBN, and 
for the layer orientation difference of particular bilayers.  Because these tight-binding model parameters
are smooth functions of $\vec{d}$, we can represent them in the moir{\' e} band 
model by a small number of parameters.  
In Figs. \ref{gbninterlayer}-\ref{gbnintralayer} we plot moir{\' e} band parameters obtained from the 
information in Fig. \ref{interlayer2} using Eq.~(\ref{integrateoverd}).
We find that the $\vec{k}$-dependence of the Hamiltonian (retained only in the $\vec{G}=(0,0)$ 
moir{\' e} band Hamiltonian term calculated by averaging the Wannier representation Hamiltonian over $\vec{d}$) 
is accurately captured by the Dirac form in both graphene and BN layers.  

The interlayer hopping physics is quite similar in the G/G and G/BN cases.
By examining Fig. \ref{interlayer2} and comparing with the previous G/G
discussion, we conclude that as for G/BN three Fourier coefficients dominate and 
yield a simple transparent model.  Including only these coefficients, we obtain 
for G/BN  
\begin{equation} 
{ H}^{bt}_{s,s'}(\vec{r}) =     \sum_{j} \,  \exp( - i \vec{Q}_j \cdot \vec{r} ) \, T^{j}_{s,s'}
\end{equation} 
where 
\begin{equation}
\label{gbntunneling}
 T^{j}=     \exp(-i \vec{G}_j  \vec{\tau})
\left(\begin{matrix} t_{BC} & t_{BC} \exp(-i j\phi) \\ t_{NC} \exp( i j\phi ) & t_{NC} \end{matrix}\right).
\end{equation}
In this case there are two distinct interlayer tunneling parameters which 
have the values $t_{BC} = 144 $ meV and $t_{NC} = 97$ meV.   
The notation is suggested by comparing these moir{\' e} band matrix elements with those
constructed from {\em ad hoc} microscopic tight-binding models. \cite{Arnaud} 
The difference between the boron to carbon and nitrogen to carbon hopping 
parameters, $t_{BC}$ and $t_{NC}$, is not unexpected since $p_z$ orbitals centered on the boron sites
should have larger atomic radii than $p_z$ orbitals centered on the larger $Z$ nitrogen sites.  
The remaining large contributions to the G/BN moir{\' e} band model
are absent for G/G and are discussed below.

In the G/BN case, coupling between layers is responsible not only 
for interlayer tunneling but also for substantial changes within the individual layers.   
From our microscopic calculations we find that the carbon site energy parameter 
is large for the first shell of reciprocal lattice vectors. 
The momentum space pattern is clearly quite different from that of the inter-layer hopping processes.
The intralayer Hamiltonian matrix elements can be presented using 
the same formulas as in Eq.~(\ref{gbn_intra}) used earlier for the G/G case.
What is different in the G/BN case is that our moir{\' e} band model is specified by two interlayer tunneling 
parameters and by $12$ intralayer parameters.
The values of the $12$ intralayer coefficients are listed below:

\begin{eqnarray}
C_{0 \,AA} &=&3.332   \,\, {\rm eV}, \quad \,\,\, C_{0 \, BB} = -1.493 \,\,{\rm eV},  \label{gbnhamiltonian}   \\
C_{0\, A'A'}  &=&   0,   \,\,  \quad \,\,\, C_{0 \,B'B'} = 0,  \nonumber \\
C_{AA}  &=&   5.733   \,\, {\rm meV}, \quad \,\,\, \varphi_{AA} = 90^{\circ},  \nonumber \\
C_{BB}  &=&   4.826   \,\, {\rm meV}, \quad \,\,\, \varphi_{BB} =  65.49^{\circ},  \nonumber  \\
C_{A'A'} &=&  -5.703  \,\, {\rm meV}, \quad \,\,\, \varphi_{A'A'} = 87.51^{\circ}, \nonumber  \\
C_{B'B'} &=&  -3.596   \,\, {\rm meV}, \quad \,\,\, \varphi_{B'B'} = 65.06^{\circ},  \nonumber  \\
C_{AB}  &=&  4.418 \,\, {\rm meV}, \quad \,\, \varphi_{AB} = 26.10^{\circ},  \nonumber \\
C_{A'B'} &=&  1.987 \,\, {\rm meV}, \quad \,\, \varphi_{A'B'} = 3.50^{\circ}.      \nonumber
\end{eqnarray}
The Fourier expansion coefficients of the Hamiltonian can be related with the above set of parameters
through the same Eqs. (\ref{relationcoefdiag}, \ref{relationcoefoffdiag}) used in the graphene on graphene case.

\subsubsection{First shell approximation for commensurate AA, AB and BA limits}

In the following we apply our model Hamiltonian to crystalline AA, AB and BA stacking limits
using the same lattice constant for both graphene and hBN. 
We proceed with our analysis in a manner similar to the G/G case.
The first shell approximation for the commensurate interlayer coupling Hamiltonian 
closely follows the G/G case, except that it is necessary to distinguish the parameters for 
tunneling from the boron site 
and the nitrogen atom site. From Eq.~(\ref{gbntunneling}) 
for the  AA ($\vec{\tau}_{AA} = (0, 0)$), AB ($\vec{\tau}_{AB} = (0, a/\sqrt{3})$) and BA ($\vec{\tau}_{AB} = (0, 2a/\sqrt{3})$)
stacking configurations we obtain: 
\begin{eqnarray}
H_{bt}(\vec{K}:\vec{\tau}_{AA}) &=&  
3 
\begin{pmatrix}
t_{BC} & 0 \\
0 & t_{NC}
\end{pmatrix},  \\     \nonumber
H_{bt}(\vec{K}:\vec{\tau}_{AB}) &=& 
3 
\begin{pmatrix}
0 & 0  \\
t_{NC} & 0
\end{pmatrix}, \\    \nonumber
H_{bt}(\vec{K}:\vec{\tau}_{BA}) &=& 
3 
\begin{pmatrix}
0 & t_{BC} \\
0 & 0
\end{pmatrix}.
\end{eqnarray}
In the first shell approximation the tunneling amplitudes are 3$t_{BC} =  432$ meV and 3$t_{NC} =  291$ meV respectively.
In comparison, direct calculations for these stacking configurations 
give  
$H_{AA^{\prime}}(\vec{K} : \vec{\tau}_{AA}) = 437 $ meV,
$H_{BB^{\prime}}(\vec{K} : \vec{\tau}_{AA}) = 294 $ meV,
$H_{BA^{\prime}}(\vec{K} : \vec{\tau}_{AB}) = 296 $ meV,
and $ H_{AB^{\prime}}(\vec{K} : \vec{\tau}_{BA}) =  439$ meV.
The deviations from 3$t_{BC}$ and 3$t_{NC}$ are in the order of a few meV which imply relative
differences smaller than 2\% for the main tunneling terms.

\subsection{Effective low energy model for G/BN} 
\label{sec:twoband} 

\begin{figure*}
\begin{center}
\includegraphics[width=18cm]{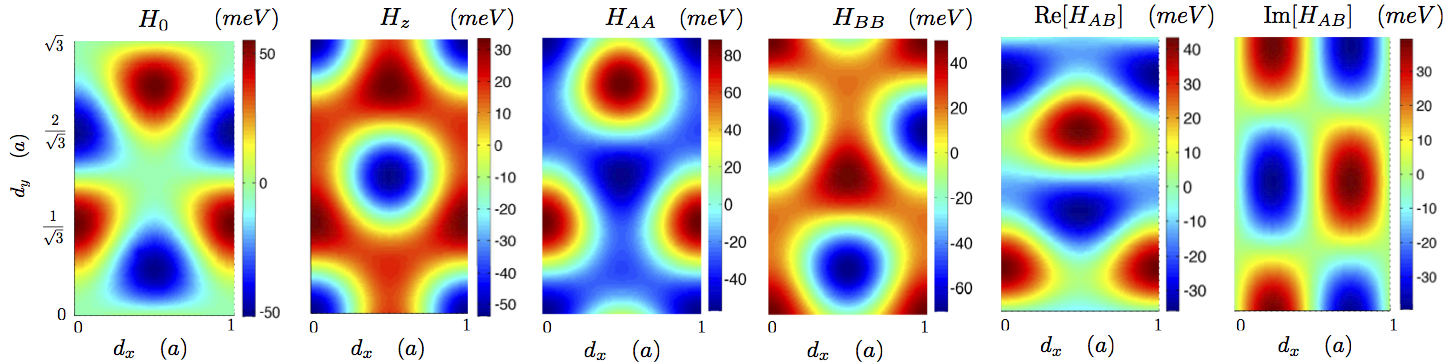} 
\end{center}
\caption{
(Color online)
Relative displacement $\vec{d}$-dependence of the matrix elements of the
two-band low-energy effective model for graphene on a hBN substrate.
In the effective model map the $H_0 = (H_{AA}+H_{BB})/2$ term represents a sublattice independent potential 
and $H_z = (H_{AA}-H_{BB})/2$ acts as a mass term in the Dirac equation. The off diagonal
matrix-elements $H_{AB}$ accounts for changes in the bonding pattern within the graphene layer. 
Here $A, B$ refer to the sublattice sites of graphene.
}
\label{gbneff}
\end{figure*}

\begin{figure}
\begin{center}
\includegraphics[width=8cm]{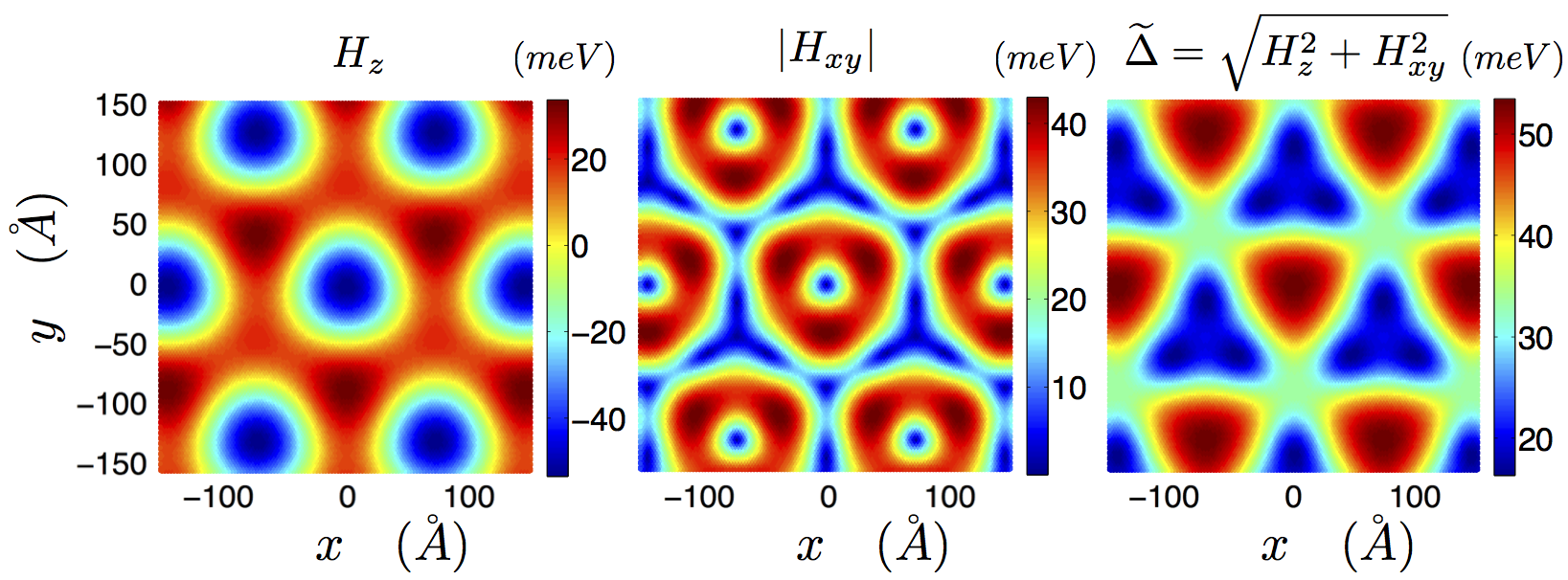} 
\end{center}
\caption{ 
Mass and pseudospin field terms in the effective Hamiltonian as a function of displacement $\vec{d}$.
The $H_z$ term is due to sublattice potential difference and vanishes along lines of this two-dimensional plot. 
The simultaneous presence of finite $H_{x}$ and $H_{y}$ and $H_z$ terms in the 
effective Hamiltonian implies that the Dirac-point gap (Eq. (\ref{diracpointgap})) 
does not vanish at any relative displacement. 
}
\label{map}
\end{figure}

\begin{figure}[htbp]
\begin{center}
\includegraphics[width=8cm,angle=0]{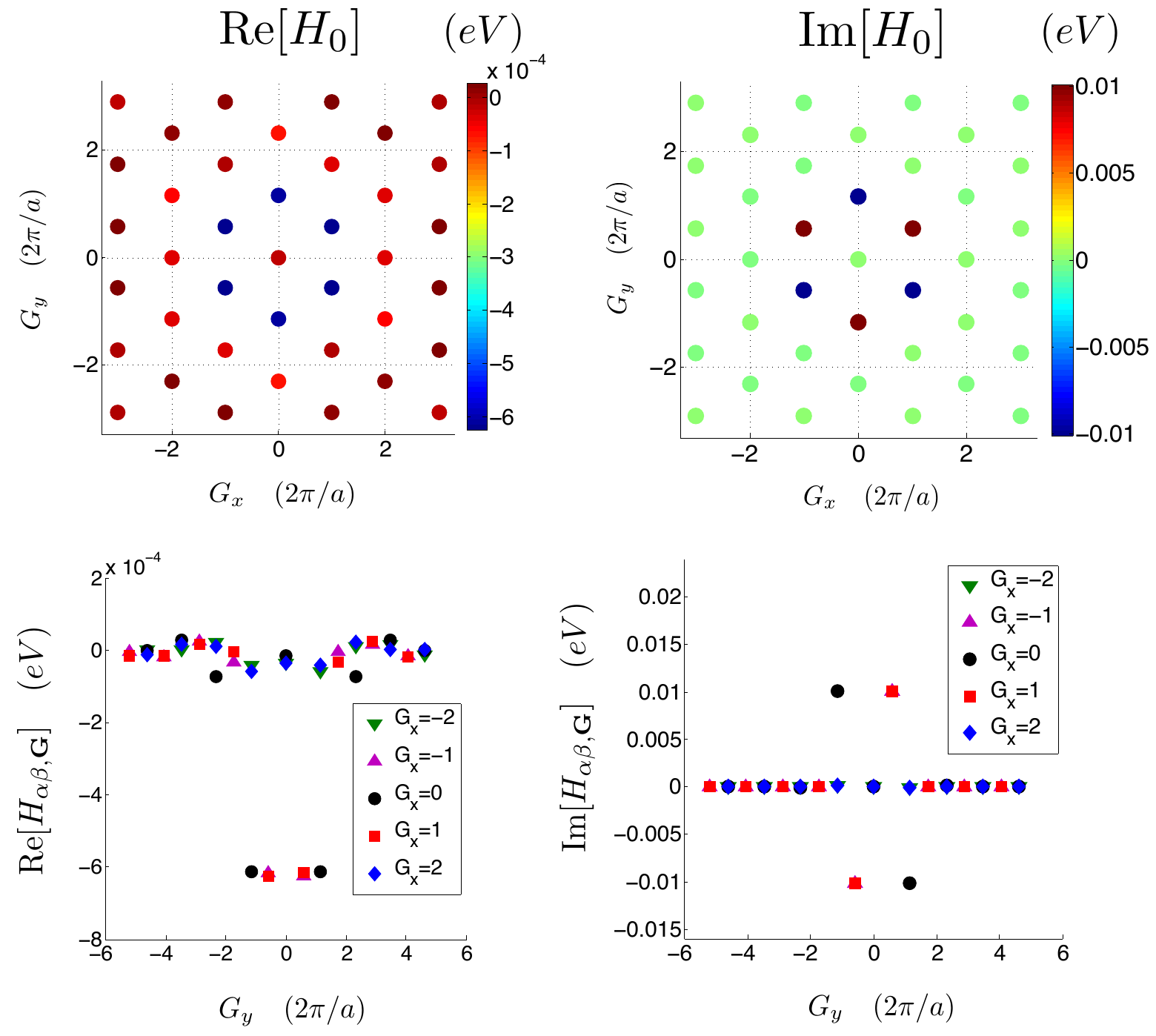}
\end{center}
\caption{
(Color online)
Fourier expansion coefficients for the effective model matrix element 
$ H_{0}(\vec{K}:\vec{d}) $.  This term captures the variation of the site-averaged potential 
across the moir{\' e} pattern.  The first shell of reciprocal lattice vectors dominates.}
\label{heff0}
\end{figure}

\begin{figure}[htbp]
\begin{center}
\includegraphics[width=8cm,angle=0]{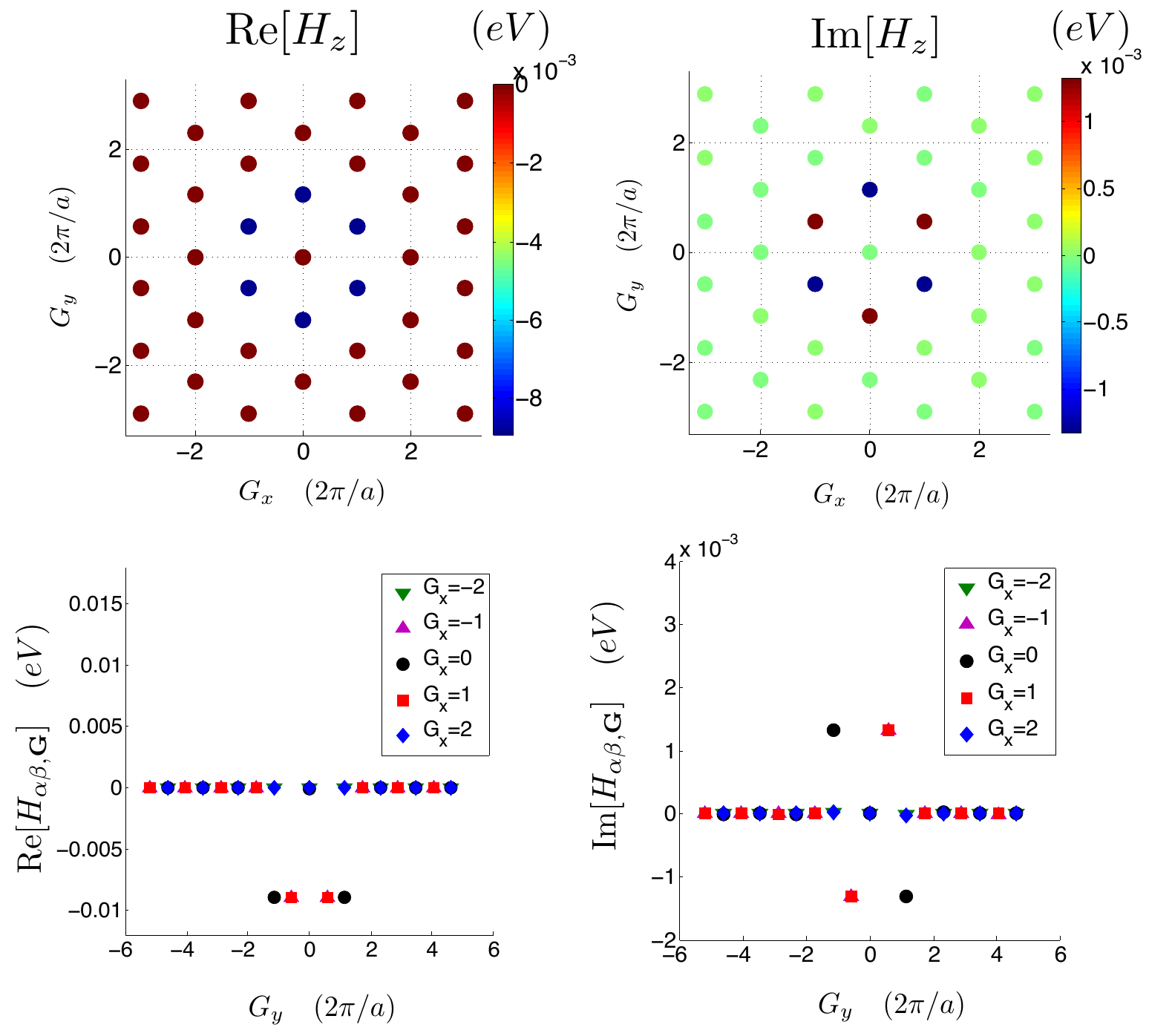}
\end{center}
\caption{
(Color online)
Fourier expansion coefficients for the mass term in the effective model,  
$H_{z} (\vec{K}:\vec{d}) = (H_{AA} (\vec{K}:\vec{d})  - H_{BB} (\vec{K}:\vec{d}) )/2$  
The first shell of reciprocal lattice vectors dominates.}
\label{heffz}
\end{figure}

Our model for the electronic structure of graphene on hBN can be further 
simplified by formulating a version which acts only on the low-energy 
degrees of freedom within the carbon layers.  We expect that this version 
of our model will be broadly applicable to describe electronic properties of graphene sheets 
that are weakly influenced by a hBN substrate.  As we see, the influence will tend to 
be stronger when the orientation angle difference between graphene and hBN layers
is small.  In this approach we integrate out the boron nitride layer degrees-of-freedom to 
obtain a two-band model for graphene. 
 
When written in terms of $2 \times 2$ blocks, 
the four-band model is given for each $\vec{d}$  by 
\begin{equation} 
H_{full} = 
\left(\begin{matrix} { H}_{BN} & { T}_{BN,G} \\
{ T}_{G,BN} & { H}_{G} \end{matrix} \right) 
\end{equation} 
where the entries in this matrix are $2 \times 2$ matrices that map sub lattices to sub lattices.
We choose the zero of energy at the carbon site energies of the graphene layer.
The effective Hamiltonian for graphene obtained by integrating out the boron nitride orbitals is 
\begin{equation} 
{ H} = { H}_{G} - { T}_{G,BN} \, { H}_{BN}^{-1} { T}_{BN,G}.
\label{perturbation}
\end{equation} 
This expression is valid to leading order in an expansion in powers of the ratio
of interlayer tunneling amplitudes to the hBN gap $\sim t_{BN}/(C_{0AA}-C_{0BB})$.  

In this two-band model we can identify four different physical effects of the hBN substrate: 
i)~There is a $\vec{d}$-dependent difference between the two carbon site energies in the honeycomb unit cell
that is absent for an isolated layer.  When viewed as a substrate contribution to graphene's two-dimensional 
Dirac equation $H_z = ( { H}_{A'A'} - { H}_{B'B'} )/2$ can be viewed as a $\vec{d}$-dependent mass.   
Note that the effective mass is sometimes positive and sometimes negative. 
The A site energy is maximized at AB points, where the carbon A site is on top of a boron atom 
and far from nitrogen atoms.  
Similarly the B site energy is minimized at BA points, where the carbon B site is  
on top of a nitrogen atom and far from boron atoms.   
ii)~$H_0 = ({ H}_{A'A'} + H_{B'B'})/2$ can be viewed as a $\vec{d}$-dependent potential term. 
iii)~$H_{A'B'}= H_{B'A'}^*$ captures the influence of the substrate on
hopping between carbon sublattices.  This quantity vanishes by symmetry at the 
Dirac point for an isolated sheet.  Our calculations demonstrate that the 
reduction in symmetry due to the substrate yields a $\vec{d}$-dependent contribution to the Hamiltonian that 
is roughly of the same size as the mass and potential terms.   
When the operators that act on sub lattice degrees of freedom are described using 
Pauli spin matrices, the real part of $H_{A'B'}$ is proportional to the coefficient of $\sigma_{x}$ 
while the imaginary part is proportional to the coefficient of $\sigma_{y}$.  
iv)~The final contribution to the effective model is due to virtual occupation of hBN sites 
and captured by the second term on the right hand side of Eq.~(\ref{perturbation}).  
The full effective Hamiltonian can be expanded in terms of Pauli matrices to yield  
an intuitive representation of the Hamiltonian's sublattice dependence.  
The term proportional to the identity matrix can be viewed as a potential term,  
the term proportional to $\sigma_{z}$ as a mass term, and 
the terms proportional to $\sigma_{x}$ and $\sigma_{y}$ as gauge 
potentials which account for substrate-induced bonding distortions. 
Virtual processes contribute to all the effective-model matrix
elements discussed above.   

The microscopic origin of the mass term can be traced
to the difference in electronegativity between nitrogen and boron which both leads
to differences in charging, and modifies the in-plane sigma bonds. 
Both effects lead to a mass term in the Hamiltonian that is proportional to $\sigma_z$.
The nitrogen (boron) is negatively (positively) charged.
Because the interlayer distance is large, 
one can crudely approximate the resulting Hartree potential 
by a Coulomb potential with an effective charge $Ze$ ($-Ze$) with $0 , Z < 1$
acting on the carbon atom just on top of it. 
This picture has been explored from a phenomenological point of view\cite{Arnaud} 
and gives rise to a mass contribution to the Hamiltonian which is qualitatively similar to
the one derived here from first principles.   

We construct the moir{\' e} band Hamiltonian by letting $\vec{d} \to \vec{d}(\vec{L})$ as 
explained in section \ref{sec:formal}.
The moir{\' e} band model is particularly simple when constructed from the 
two-band effective model:  
\begin{equation} 
H_{ss'} = H^{0}_{ss'} + H^{MB}_{ss'} 
\label{effham1}
\end{equation} 
where $H^{0}_{ss'}$ is the non-local Hamiltonian which describes the Dirac cones,
\begin{equation} 
H^0_{ss'} = H_{s,s'}(\vec{k}:\vec{G}=0)  \delta_{\vec{k},\vec{k'}}  
\end{equation} 
and $H^{MB}_{ss'}$ is the term which captures the moir{\' e} band 
modulation:
\begin{equation} 
H^{MB}_{ss'} = \sum_{\vec{G}\ne0}  H_{s,s'}(\vec{K}:\vec{G})   
\; \Delta(\vec{k'}-\vec{k}  -\tilde{G}) .  
\label{matrixelementb}
\end{equation}
This model can be viewed as the Hamiltonian of graphene subject to external periodic pseudospin-dependent
potentials represented in a Fourier expanded form as a sum in the $\widetilde{G}$ lattice vectors of the moir{\' e} 
reciprocal lattice.
The form of the Hamiltonian is informed by 
first principles calculations that account
not only for the variation in carbon layer site-energies with local coordination, 
but also for variations in inter-carbon hopping and for virtual hopping between graphene and boron nitride layers.
As we will show shortly, thanks to the smooth displacement dependence of the $\vec{d}$-dependent 
effective Hamiltonian, the moir{\' e} patterns of the pseudospin fields are 
accurately captured by three pairs of parameters, 
one pair for each pseudospin effective field component.  
These generalized superlattice potentials
determine the quasiparticle velocity and gaps  
in the moir{\' e} superlattice band structure.\cite{superlatticepotentials}  
Our model provides a simple and
accurate starting point from which we can calculate the electronic structure of 
graphene superlattices subject to moir{\' e} patterns of the 
pseudospin fields shown in Figs. \ref{gbneff} and \ref{map}. 

\begin{figure}[htbp]
\begin{center}
\includegraphics[width=8cm,angle=0]{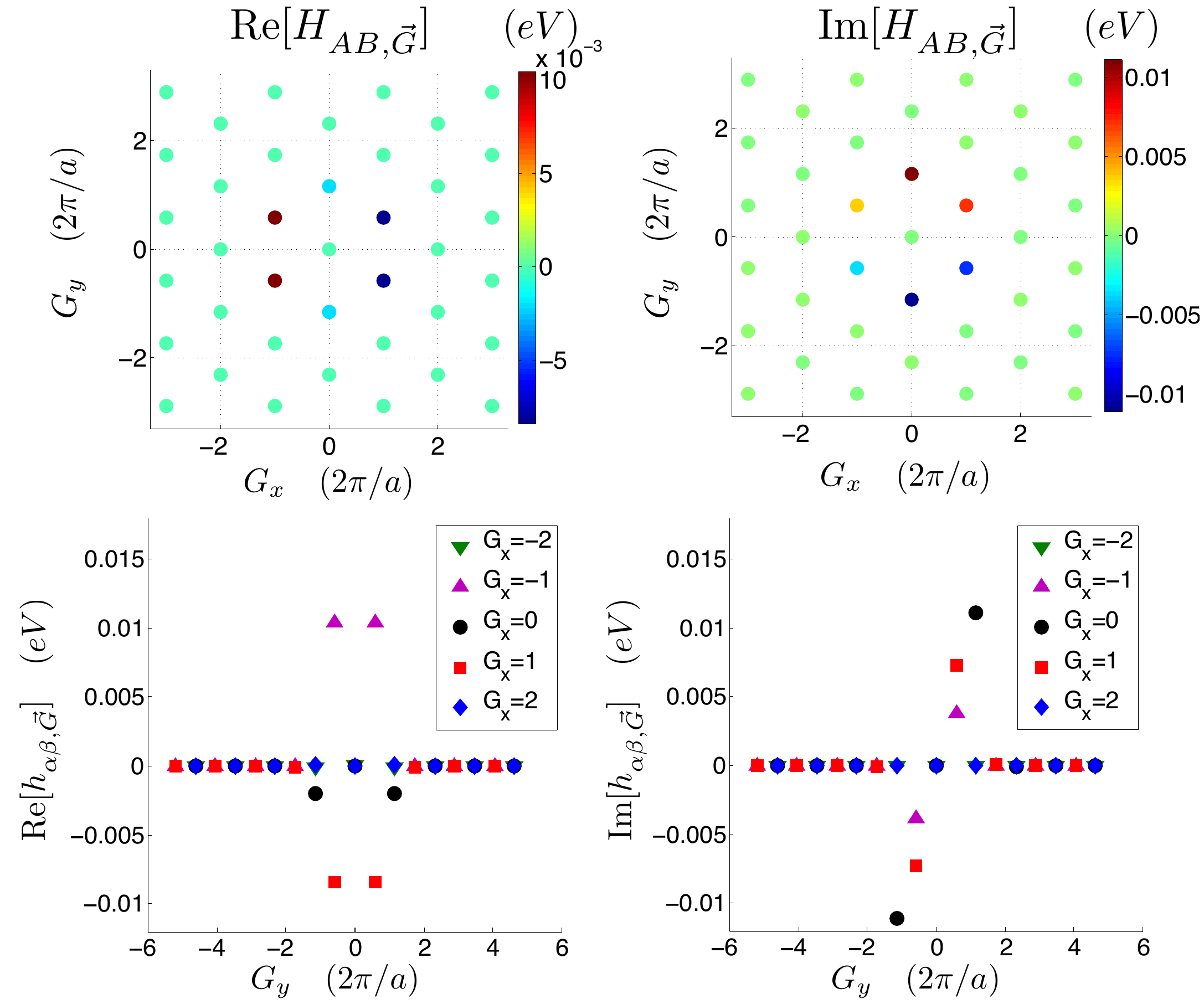}
\end{center}
\caption{
(Color online)
Fourier expansion coefficients
for the off diagonal effective model matrix element 
$  H_{AB} (\vec{K}:\vec{d})$. The first shell of reciprocal lattice vectors dominates.
}
\label{heffab}
\end{figure}

Our numerical results for the Fourier expansion coefficients of the effective model matrix elements are 
summarized in Figs. \ref{heff0}-\ref{heffab}.
Once again the expansion coefficients are dominated by the first 
shell of $\vec{G}$'s, and the number of independent coefficients is reduced by symmetry.  
We find that  
\begin{widetext}
\begin{eqnarray}
H_{0}(\vec{K} :\vec{d}) &=&  2 C_{0} {\rm Re}[ f(\vec{d})  \exp[i \varphi_{0}]  ] , \quad \quad 
H_{z}(\vec{K} :\vec{d}) =  2 C_{z} {\rm Re}[ f(\vec{d})  \exp[i \varphi_{z}]  ]      \label{hofd} \\
H_{AB}(\vec{K} :\vec{d}) &=& 2 C_{AB} \, \cos( \frac{\sqrt{3} }{2} G_1 d_x) 
\left(  
\cos \left( \frac{G_1 d_y}{2}  - \varphi_{AB}  \right) 
+\sin \left( \frac{G_1 d_y}{2}  - \varphi_{AB} - \frac{\pi}{6}  \right)
\right)  
+ 2 C_{AB} \, \sin \left( G_1 d_y + \varphi_{AB} - \frac{\pi}{6}  \right)   \nonumber \\
&+& i \,2 C_{AB} \, \sin( \frac{\sqrt{3}  }{2} G_1 d_x ) 
\left(  
\cos \left( \frac{G_1 d_y}{2}  - \varphi_{AB}  \right)   
- \sin \left( \frac{G_1 d_y}{2}  - \varphi_{AB} - \frac{\pi}{6}  \right)
\right)    
 \nonumber 
\end{eqnarray}.
\end{widetext}
The site-independent term $H_0$ gives rise to an overall potential shift in the 
graphene layer depending on the local stacking order.
The pseudospin in-plane terms $H_x$ and $H_y$ together with  and the mass term $H_z$, 
are the coefficients of 
$\sigma_x$, $\sigma_y$, and $\sigma_z$ in the local $2 \times 2$ moir{\' e} band Hamiltonian.
The in-plane pseudospin terms $H_x$ and $H_y$, can be viewed as a gauge fields $\vec{A}$, \cite{vozmediano}
that shift the Dirac cone band edges away from the original position.
Together these coefficients determine the local Dirac point gap through the relation
\begin{equation} 
\label{diracpointgap}
\widetilde{\Delta}(\vec{r}) = 2 \; \sqrt{H_x^2(\vec{r})+ H_y^2(\vec{r}) + H_z^2(\vec{r})}.
\end{equation} 
This local Dirac point gap is not directly related to the overall gap of the 
moir{\' e} pattern because of the non-locality of the momentum-dependent 
isolated layer Dirac Hamiltonian.  (We also expect that the gap will
be strongly influenced by many-body effects.) 
We see in Fig. \ref{map} that the Dirac point gap is everywhere 
at least $30$ meV because $H_x$, $H_y$ and $H_z$ do not vanish simultaneously. 

Our effective model is completely specified by six numbers:  
\begin{eqnarray}
C_{0} &=&  -10.13 \,\, {\rm meV}, \quad \,\,\,   \varphi_{0} = 86.53^{\circ},  \label{parameters}    \\ 
C_{z} &=&   -9. 01\,\, {\rm meV}, \quad  \,\,    \varphi_{z} = 8.43^{\circ},   \nonumber  \\ 
C_{AB} &=& \,\,\, 11.34 \,\, {\rm meV}, \quad   \,\,\,\, \,\,   \varphi_{AB} = 19.60^{\circ}.    \nonumber 
\end{eqnarray}

\begin{figure} 
\begin{tabular}{c}
\includegraphics[width=8.4cm,angle=0]{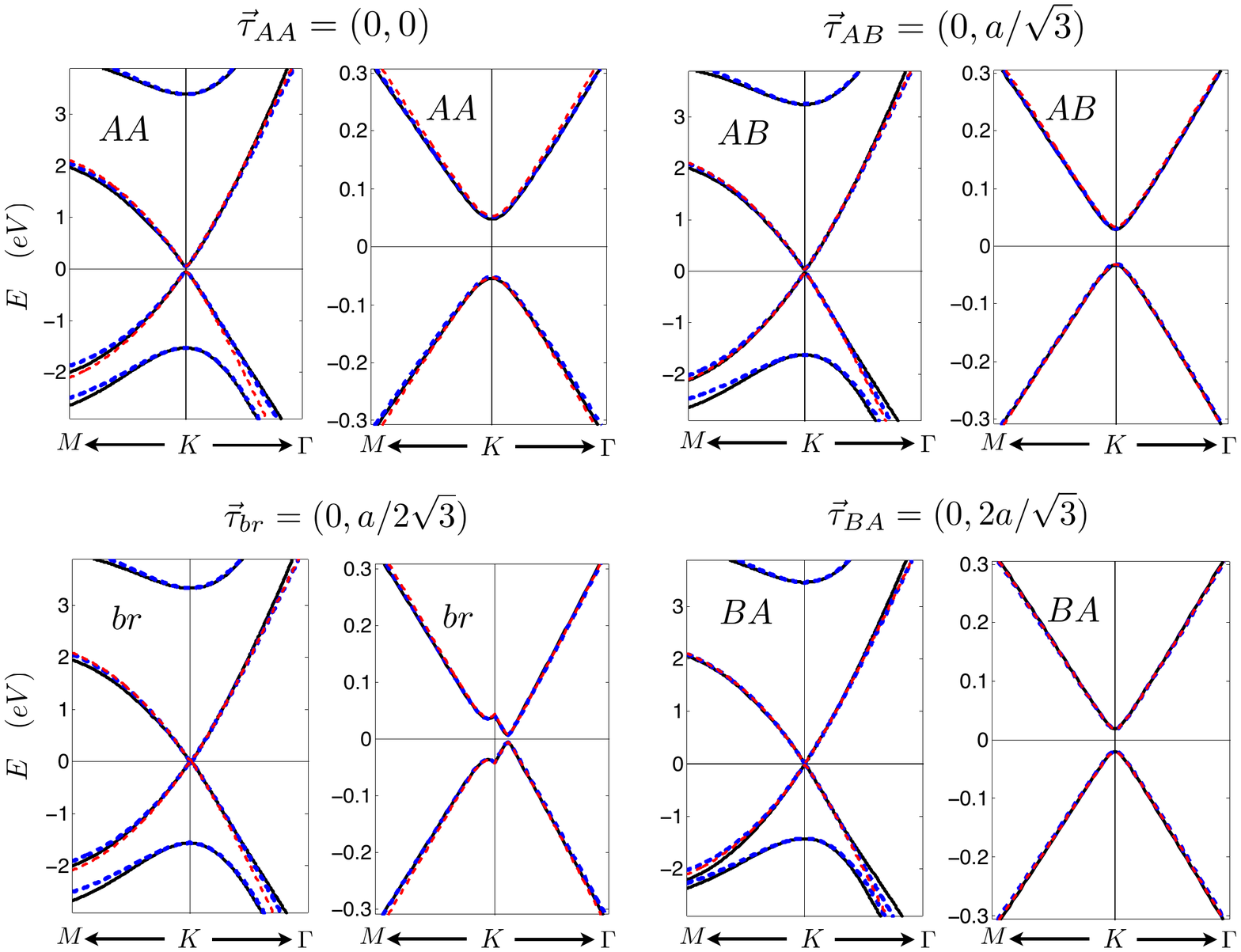} 
\end{tabular}
\caption{
(Color online) 
Comparison of the LDA band structure (solid black), the four-band moir{\' e} band model (dashed blue lines) 
and the low energy two-band model (dashed red lines) with the first shell used for the superlattice potentials. 
The commensurate G/BN arrangements plotted are 
AA, AB, BA and a bridge stacking with $\vec{\tau}_{br}=(0, a/2\sqrt{3})$.
Note that he electronic structures of AB and BA stacking are different in the 
G/B case. For the intermediate bridge stacking we see a substantial reduction of the band gap due to a  
shifting in the Dirac cone momentum space location caused by in-plane pseudospin terms. 
The intralayer Hamiltonian of graphene is approximated using the massless Dirac model with the LDA Fermi velocity
\cite{graphenetb} while the boron nitride Hamiltonian is modeled with the same Dirac model with
a mass term compatible with the LDA gaps. The interlayer coupling is given by the first shell approximation using 
the parametrizations of Eq.~(\ref{gginterlayer}), with tunneling from boron and carbon sites
distinguished as in Eq.~(\ref{gbntunneling}) and the parameter
set in Eq.~(\ref{gbnhamiltonian}).
In these plots the energy origin of the represented bands has
been adjusted so that zero is in the middle of the band gap.
}
\label{gbcommensurate}
\end{figure}

As described in detail for the G/G case, wave vector reduced to the moir{\' e}
Brillouin-zone is a good quantum number for this model, and band eigenstates may
be obtained by making plane-wave expansions.  The graphene layer  
Dirac Hamiltonian contributes to diagonal blocks in the plane-wave 
representation of the moir{\' e} band Hamiltonian.
The Fourier expansion of the Hamiltonian in $\vec{G}$ vectors can be related to the above parameters through
Eqs. (\ref{relationcoefdiag}) for the diagonal terms, either in the pseudospin or sublattice basis, 
and for the off-diagonal terms shown in Fig.~\ref{heffab} we have the following form
\begin{eqnarray}
\label{relationcoefeff}
H_{AB, \vec{G}_1} &=& H^{*}_{AB, \vec{G}_4} = C_{AB} \exp(i (  2 \pi / 3  - \varphi_{AB} )),    \\
H_{AB, \vec{G}_3} &=& H^{*}_{AB, \vec{G}_2} = C_{AB} \exp(-i  \varphi_{AB} ),   \nonumber \\
H_{AB, \vec{G}_5} &=& H^{*}_{AB, \vec{G}_6} = C_{AB} \exp(  i(  -2 \pi / 3  - \varphi_{AB} )).  \nonumber
\end{eqnarray}

\begin{figure}
\begin{center}
\includegraphics[width=8cm]{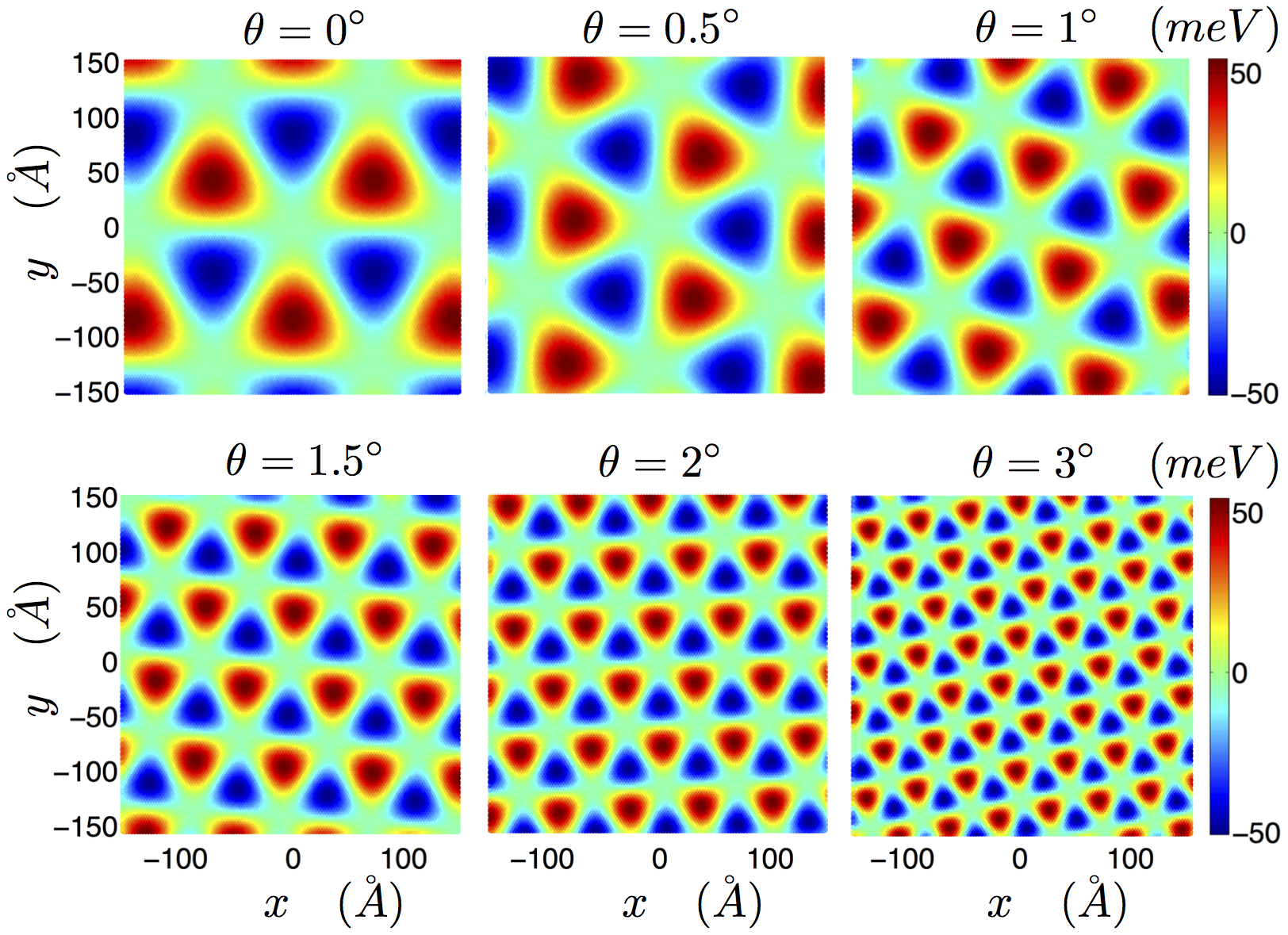} 
\end{center}
\caption{ 
Modulation of the local potential fluctuations $H_{0}(\vec{r})$ in real space for different twist angles. 
These plots illustrate the rotation of the moire pattern when $\left| \varepsilon \right| \sim \theta$,
and the property that the Moire periodicity $L_{M} \sim a/\sqrt{\varepsilon^2 + \theta^2}$ becomes shorter with increasing twist angle.
The other pseudospin components of the local Hamiltonian illustrated Fig. \ref{gbneff} produce
similar spatial superlattice patterns.
}
\label{v0map}
\end{figure}

The applicability of the effective model is evidenced by its accuracy in describing the 
band structure for the commensurate stacking arrangements shown in Fig. \ref{gbcommensurate}.
In these plots local potential fluctuations due to $H_0$ are manifested by a small offset between the graphene and hBN bands.
In the presence of a finite twist angle, the $H_0$ term leads to an effective potential that varies in 
space as shown in Fig. \ref{v0map}.
These potentials variation on the Moire pattern scale leads to 
the local density-of-state variations seen experimentally. \cite{LeRoy} 

\begin{figure}
\begin{center}
\includegraphics[width=8cm]{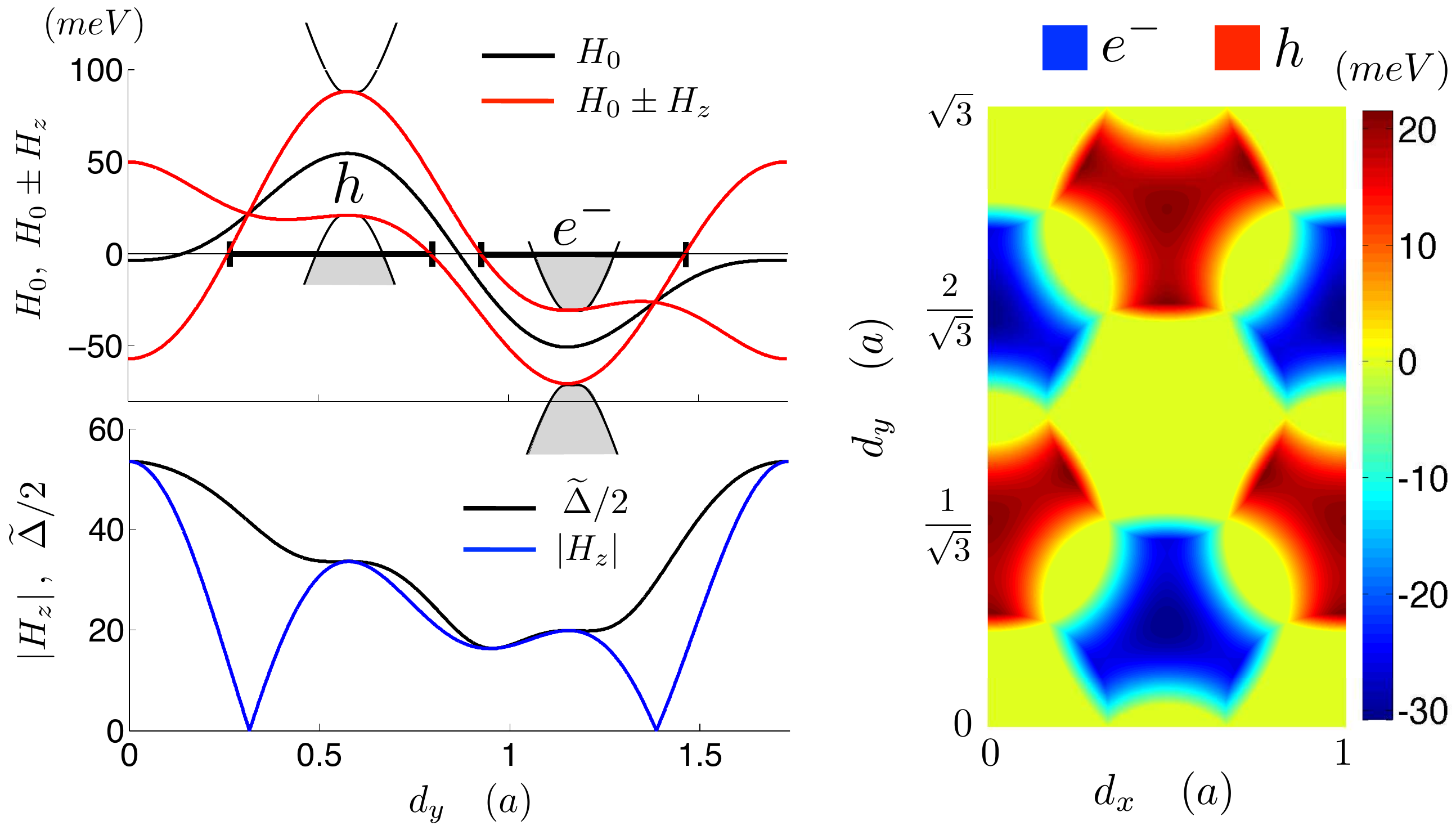} 
\end{center}
\caption{ (Color online)
{\em Left Panel:}
Schematic illustration of potential variations and local band edges as a function of $d_y$ 
for fixed $d_x=0$. The approximate conditions for local electron/hole charging
discussed in the main text are satisfied over the segments identified by bold black lines.
The difference between the Dirac point gap $\widetilde{\Delta}$ and the absolute value of the mass $\left| H_z \right|$
reflects the influence of the in-plane pseudospin fields terms.  
{\em Right Panel:}
Sliding vector $\vec{d}$ dependent map of electron and hole puddles. 
}
\label{ehpuddles}
\end{figure}

Even when a graphene sheet on a hBN substrate is globally neutral the charge 
density will vary locally.  The regions within the moir{\' e} pattern
in which positive and negative charge densities are expected can be identified by neglecting the 
non-local Dirac Hamiltonian (which vanishes at the Dirac point) and the $H_{x}$ and $H_{y}$ sublattice coupling terms.
In this limit charge puddles should be expected wherever the chemical potential, set by imposing global charge neutrality,
lies below the lower sublattice site energy or above the upper sub lattice site energy.  
Since the chemical potential at neutrality is very close to the average site energy,
which we have chosen as the energy zero, this condition for the formation of charge puddles
is equivalent to $\left|H_0(\vec{r})\right| > \left| H_z (\vec{r})\right|$, 
with the carrier type being electrons if $H_0 > H_z$ and holes if $H_0<H_z$. 
In Fig. \ref{ehpuddles} we apply this criterion to obtain a map of electron and hole puddles from
the parametrization for $H_0$ and $H_z$ presented in Eq. (\ref{hofd}). 
To obtain a more quantitatively accurate map it will be necessary to 
restore the $H_{x}$ and $H_{y}$ and to take into account other effects such as
the electrostatic screening and many-body corrections. 
Lattice relaxations, whose influence is discussed in the appendix, will
also play a role.  

\section{Summary and Discussion} 
\label{sec:discussion} 

We have presented a method which can be used to derive approximate electronic structure 
models for layered semiconductors, semimetals, and gapless semiconductors
containing a finite number of two-dimensional crystals with
different lattice constants and/or different crystal orientations.  
The method is intended to be useful for multilayer graphene systems,
multilayer transition metal dichalcogenide systems, and for multi-layer systems containing 
both graphene and boron nitride. 
When several layers are present simultaneously, structures of this type are 
not in general two-dimensional crystals, and electronic structure theory
can therefore be awkward to apply directly because Bloch's theorem is not valid.

Our approach focuses on the influence on the electronic 
structure of slowly varying relative displacements $\vec{d}(\vec{L})$ between
individual crystalline layers due to a small difference in lattice
constants or crystal orientations.  The dependence of electronic structure on $\vec{d}$ 
can be calculated without experimental input using density-functional-theory.
Our analysis produces a moir{\' e} band model which is periodic under translations $\vec{R}$ for which 
\begin{equation} 
\vec{d}(\vec{L}+\vec{R})=\vec{d}(\vec{L})+\vec{L}'
\label{moirecondition}
\end{equation} 
for some two-dimensional lattice
vector $\vec{L}'$.  
The system is microscopically crystalline
only if the vectors $\vec{R}$ for which Eq.~(\ref{moirecondition}) is satisfied are 
lattice vectors of the two-dimensional crystal.
The vectors $\vec{R}$ are the lattice vectors of the moir{\' e} pattern. 
Like the moir{\' e} pattern itself,\cite{Amidror09}
our moir{\' e} band models have a periodicity defined by spatially varying layer alignment, 
and can therefore be analyzed using  
Bloch's theorem for a superlattice Hamiltonian that has the periodicity of the moir{\' e} pattern.  
The models consists of massless or massive Dirac models for each two-dimensional layer,
combined with a spatially local effective potential which acts on sublattice degrees of freedom.  

Our approach to coupled bilayer systems has three main limitations.  First of all it does not 
apply to cases in which differences in lattice constants or rotation angles between 
adjacent layers are large.
This limitation can be overcome, however, by building a theory that 
is based on larger unit cells with more sub lattice sites and lattice constant ratios
between neighboring layers closer to one.  For example, for G/G one could for example build models
that are similar to the ones discussed here which would apply at rotation angles close to the 
short-period commensurate rotation angles.   
Secondly, because it attempts to describe bands over 
a relatively small part of the Brillouin-zone, it is valid over a limited energy range.
Finally, it assumes that the individual layers are indeed crystalline whereas we should in fact  
expect that the moir{\' e} pattern will induce small structural distortions within each layer.
For the van der Waals epitaxial systems of interest, however, it seems reasonable to 
expect these distortions to be small and to neglect them, at least as a first approximation. 

We have applied our moir{\' e} band method to two different two-layer systems, one with 
two graphene layers and one with a graphene layer and a hexagonal boron nitride layer. 
For the case of graphene on boron nitride, which has a large energy gap, 
we have also derived a simpler model, specified by Eqs.~(\ref{effham1},~\ref{parameters}),
in which the boron nitride degrees of 
freedom are treated perturbatively to obtain an explicit model for graphene on a 
boron nitride substrate which retains only the 
graphene $\pi$-electron degrees of freedom.
In the case of graphene on graphene our 
calculations explain why the dependence on relative orientation angle of 
G/G electronic structure is accurately described by a model with a single-interlayer 
tunneling parameter.  For the case of graphene on boron nitride, the models we produce are  
more complicated because of the need to account for the dependence on 
$\vec{d}$ of graphene-layer site energies and interlayer tunneling amplitudes,
but still have a small number of parameters.  Nevertheless, the graphene only model for 
G/BN is able to accurately describe the dependence of low-energy bands on rotation 
angle using six parameters which we have calculated from the $\vec{d}$-dependence of 
{\em ab initio} bands.     

The models derived in this 
paper can be used as a starting point to account for the influence of either graphene or hBN 
substrates on the electronic structure of a graphene layer.  We expect that 
they will be applicable to examine a wide variety of electronic properties.  
Because our models are derived from local-density-approximation band-structures, they 
do not account for the non-local exchange and correlation effects which are known 
to be responsible for large Fermi velocity enhancements in isolated graphene systems.\cite{nonlocal}
The same effects are likely to be important in multi-layer systems, possibly enhancing 
band gaps produced by the moir{\' e} pattern potentials.  \cite{levitov}   
Our electronic structure models are sufficiently simple that 
important many-body physics effects can be addressed separately where 
they play an essential role.

\begin{acknowledgments}
This work was supported by the Department of Energy, Office of Basic Energy Sciences under contract DE-FG02-ER45118, 
and by Welch Foundation grant TBF1473. 
JJ was partially supported by the National Research Foundation of Singapore under its Fellowship program (NRF-NRFF2012-01).
Helpful conversations with Rafi Bistritzer and Byounghak Lee are gratefully acknowledged. 
We appreciate assistance and computer time provided by  
the Texas Advanced Computing Center. 
\end{acknowledgments}

\begin{appendix}

\section{Exact form of the displacement vector and moire reciprocal lattices}
\label{exactform}

In the main text we used the small angle approximation for the displacement vectors
and the moire reciprocal lattices.  These approximations are 
accurate for rotations angles $\theta \simeq 15^{\circ}$.
Here we present the exact form of the lattice vector scaling and rotation transformation
and the moire reciprocal lattice vectors.
Let us consider the rotation operator 
\begin{eqnarray}
{\cal R}(\theta) = 
\begin{pmatrix}
\cos \theta & - \sin \theta \\
\sin \theta & \cos \theta
\end{pmatrix}
\end{eqnarray}
and write the local displacement vector as
\begin{eqnarray}
\vec{d}(\vec{L}) = \alpha {\cal R}(\theta) \vec{L} - \vec{L} = \left( \alpha {\cal R}(\theta) - 1 \right) \vec{L} = \widetilde {\cal R}(\alpha,\theta ) \vec{L} 
\end{eqnarray}
where we have defined a new transformation operator $\widetilde{ \cal R}(\alpha, \theta)$, 
$\alpha$ is the scaling ratio and $\theta$ is the twist angle of the top layer with respect to bottom layer.
The magnitude $\beta$ is the scaling factor associated to the transformation
\begin{eqnarray}
\left| \widetilde{R}(\alpha,\theta) \vec{L} \right| = \beta \left| \vec{L} \right|
\end{eqnarray}
where  $\beta = (  \varepsilon^2 + (1 + \varepsilon)(2 - 2 \cos \theta) )^{1/2}$ 
was approximated in the main text as $ \sim (\varepsilon^2 + \theta^2)^{1/2}$ when we neglected the higher order corrections.

The moire reciprocal lattice vectors $\widetilde{G}$ can be obtained applying the adjoint of the transformation operation ${\cal R}(\alpha, \theta)$ on
the $\vec{G}$ vectors
\begin{eqnarray}
 \begin{pmatrix}
 \widetilde{G}_x \\
 \widetilde{G}_y
 \end{pmatrix}
 &=&
\begin{pmatrix}
\alpha \cos \theta -1  & \sin \theta \\
- \alpha \sin \theta & \alpha \cos \theta -1
\end{pmatrix}   
\begin{pmatrix}
 G_x \\
 G_y
 \end{pmatrix}  \\
&=&
\beta
\begin{pmatrix}
 \cos \widetilde{\theta}  & - \sin \widetilde{\theta} \\
 \sin \widetilde{\theta} &  \cos \widetilde{\theta} 
\end{pmatrix} 
\begin{pmatrix}
 G_x \\
 G_y
 \end{pmatrix}.
\end{eqnarray}
We defined the rotation angle $\widetilde{\theta}$ of the moire reciprocal lattice vectors with respect to the original
$\vec{G}$ vectors and they can be obtained from the relations
\begin{eqnarray}
\widetilde{\theta} &=& \cos^{-1} \left(  \left( \alpha \cos \theta -1 \right) / \beta \right) \\
\widetilde{\theta} &=& \sin^{-1} \left( - \alpha \sin \theta / \beta \right).
\end{eqnarray}

\section{Influence of vertical lattice constant relaxation}
\label{relaxed}
In this appendix we present another set of model parameters obtained allowing lattice relaxation 
for the interlayer distance in the self-consistent LDA calculations, instead of fixing the vertical atomic separations 
at the experimental interlayer spacing $c = 3.35\,\, \AA$ of graphite.
Even though the LDA approximation does not accurately captures the non-local van der Waals type interlayer interactions
that are important in these systems, its tendency to over bind covalent bonds allows it to hold  the weakly interacting layers together.
and describe the interlayer lattice constants and the forces between 
van der Waals layered materials reasonably well.  
LDA results tend to have reasonable agreement with 
sophisticated RPA and beyond total energy calculations for 
thin jellium metal slabs,\cite{jelliumrpa} hexagonal boron nitride,\cite{hbnacfd} and graphite\cite{graphitebinding}
and other layered materials.\cite{bjorkman}
We allowed relaxation of the atomic positions in the out of plane
$z$ direction using the same 42$\times$42  $k$-points grid and using a slightly coarser threshold of total 
energy convergence in the geometry relaxation of $10^{-8}$ a.u. per unit cell and total force of $10^{-7}$ a.u.
In both G/G and G/BN cases the overall effect of the relaxation is to
increase the interlayer separation by $\sim$ 0.2~$\AA$ with respect to the closest interlayer separation. 
Theis changes leads to a weakening of the interlayer coupling strength in the first shell approximation by about 7\% .
which implies that farther $\vec{G}$ vectors in the Fourier expansion become more important.
The changes in the position of the Fermi energy with respect to the unrelaxed geometry are 
in the order of $\sim 10$~meV for G/G and $\sim 30$~meV for G/BN providing a measure
of changes in the shifts in the site potential offsets between the layers near the AA stacking configurations that 
are incorporated in the relaxed parameter set. 
Because the pseudopotentials are referenced to vacuum this information
gives an estimate of the modulation in the work function of the graphene sheet due to its coupling with the hBN layer.
The above observations suggest that geometry relaxation can introduce small but non-negligible changes in the
potential map and details of valence-conduction bands overlap in the G/BN case whose 
band structure near the Fermi energy is determined simultaneously by in-plane $xy$ and $z$ pseudospin terms 
of comparable magnitudes. 
This and other details of the electronic structure in a G/BN hetrostructure will be presented elsewhere.

\begin{figure}
\begin{center}
\begin{tabular}{cc}
\includegraphics[width=8.5cm,angle=0]{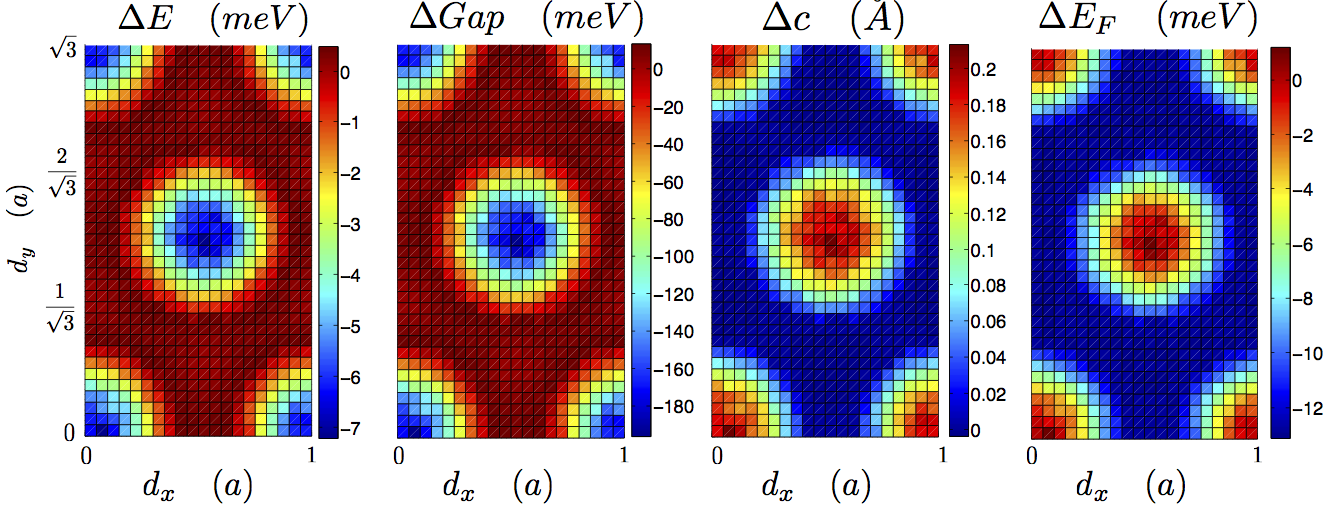}  
\end{tabular}
\end{center}
\caption{(Color online)
Changes in total energy, Dirac point gaps, average layer separation distance and Fermi energy 
resulting from allowing self-consistent LDA relaxation in the out of plane $z$-axis for G/G.
 }
 \label{ggrelaxed}
\end{figure}

\subsection{Relaxed geometry parameters for G/G}
The $\vec{d}$-vector dependent maps of the Hamiltonian matrix elements 
are changed only quantitatively relative to the unrelaxed case.
The numerical values of the parameters that define the intralayer model for G/G in Eq.~(\ref{gbn_intra}) for relaxed geometries are
\begin{eqnarray}
C_{AA} &=& 2.3 \,\, {\rm meV}, \quad \,\,\, \varphi_{AA} = 27.5^{\circ},     \\
C_{BB} &=& C_{AA} \,\, , \quad \,\,\, \varphi_{BB} = - \varphi_{AA},     \nonumber  \\
C_{AB} &=& 2.08 \,\, {\rm meV}.       \nonumber   
\end{eqnarray}
whereas the interlayer tunneling constants are $t_{bt} = 98$~meV.
The average interlayer separation distance lies between the minimum 3.347 $\AA$ and the maximum of  3.563 $\AA$ for AA stacking.

\subsection{Relaxed geometry parameters for G/BN}
\begin{figure} 
\begin{center}
\begin{tabular}{cc}
\includegraphics[width=8.5cm,angle=0]{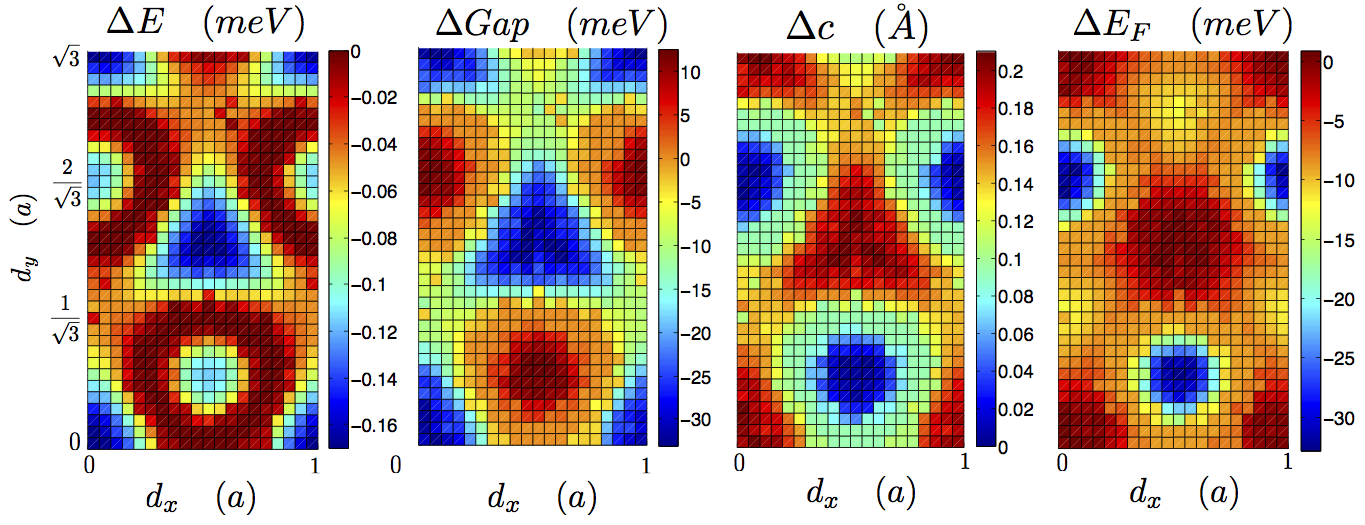}
\end{tabular}
\end{center}
\caption{(Color online)
Changes in total energy, Dirac point gaps, average layer separation distance 
measured from the minimum separation distance and Fermi energy 
resulting from allowing self-consistent LDA relaxation in the out of plane $z$-axis for G/BN.
 }
 \label{gbnrelaxed}
\end{figure}
For G/BN we have kept the coordinates fixed for the BN sheet while we allowed the carbon
atoms to relax in the out of plane direction as a function of $\vec{d}$.  T
The $\vec{d}$-vector dependent map of the Hamiltonian matrix elements for G/BN for relaxed geometries 
also has some quantitative changes relative to the unrelaxed calculations.
The numerical values of the parameters that define the intralayer model for G/BN together with the Eqs. (\ref{gbn_intra}) for 
relaxed geometries are
\begin{eqnarray}
C_{0 \,AA} &=&3.334   \,\, {\rm eV}, \quad \,\,\, C_{0 \, BB} = -1.494 \,\,{\rm eV},  \label{gbnhamiltonianrelaxed}  \\
C_{0\, A'A'}  &=&   0,   \quad \,\,  \quad \,\,\, C_{0 \,B'B'} = 0,  \nonumber \\
C_{AA}  &=&   5.643   \,\, {\rm meV}, \quad \,\,\, \varphi_{AA} = 56.37^{\circ},  \nonumber \\
C_{BB}  &=&   4.216   \,\, {\rm meV}, \quad \,\,\, \varphi_{BB} =  59.98^{\circ},  \nonumber  \\
C_{A'A'} &=&  -7.402  \,\, {\rm meV}, \quad \,\,\, \varphi_{A'A'} = 77.71^{\circ}, \nonumber  \\
C_{B'B'} &=&  -4.574   \,\, {\rm meV}, \quad \,\,\, \varphi_{B'B'} = 85.78^{\circ},  \nonumber  \\
C_{AB}  &=&   4.01 \,\, {\rm meV}, \quad \,\, \varphi_{AB} = 22.2^{\circ},  \nonumber \\
C_{A'B'} &=&  1.90 \,\, {\rm meV}, \quad \,\, \varphi_{A'B'} = 1.30^{\circ}.   \nonumber
\end{eqnarray}
whereas the interlayer tunneling constants are  $t_{BC} = 130$~meV and $t_{NC} = 87$~meV.
The average interlayer separation distance lies between 3.256 $\AA$ and the maximum of 3.466 $\AA$ for AA stacking.
The parameters of the effective model of G/BN 
for the relaxed geometries in the sublattice basis are given by
\begin{eqnarray}
C_{AA} &=& -13.3 \,\, {\rm meV}, \quad \,\,\,   \varphi_{AA} = 63.63^{\circ},   \\ 
C_{BB} &=&  14.0 \,\, {\rm meV}, \quad  \,\,    \varphi_{BB} = -51.27^{\circ},  \nonumber  \\ 
C_{AB} &=& \,\,\, 9.53 \,\, {\rm meV}, \quad   \,\,\,\, \,\,   \varphi_{AB} = 21.82^{\circ}.    \nonumber
\label{parametersrelax} 
\end{eqnarray}
The parameters in the sublattice basis and in the pseudospin basis can be related through
\begin{eqnarray}
C_{AA}  &=& - \sqrt{  C_{0}^2 + C_{z}^2 - 2 C_0 C_z \cos (\varphi_0 - \varphi_z) }    \label{equiv} \\
\varphi_{AA} &=&  \tan^{-1}\left( \frac{C_0 \sin(\varphi_0) - C_z \sin(\varphi_z)}{ C_0 \cos(\varphi_0) - C_z \cos(\varphi_z)}\right)  \nonumber   \\
C_{BB}  &=&   \sqrt{  C_{0}^2 + C_{z}^2 + 2 C_0 C_z \cos (\varphi_0 - \varphi_z) }    \nonumber    \\
\varphi_{BB} &=&  \tan^{-1} \left( \frac{C_0 \sin(\varphi_0) + C_z \sin(\varphi_z)}{  (C_0 \cos(\varphi_0) + C_z \cos(\varphi_z)} \right).    \nonumber
\end{eqnarray}
\end{appendix}

\end{document}